\documentclass[12pt,preprint]{aastex}

\newcommand{\msun}{M_\odot}
\newcommand{\rsun}{R_\odot}

\catcode`\@=11
\newcommand{\gapprox}{\mathrel{\mathpalette\@versim>}}
\newcommand{\lapprox}{\mathrel{\mathpalette\@versim<}}
\newcommand{\propapprox}{\mathrel{\mathpalette\@versim\propto}}
\newcommand{\@versim}[2]
  {\lower3.1truept\vbox{\baselineskip0pt\lineskip0.5truept
\ialign{$\m@th#1\hfil##\hfil$\crcr#2\crcr\sim\crcr}}}
\catcode`\@=12

\shorttitle{IR Observations of PWN 0540-69}
\shortauthors{WILLIAMS ET AL.}

\begin{document}

\title{Ejecta, Dust, and Synchrotron Radiation in B0540-69.3: A More
Crab-like Remnant than the Crab}

\author{Brian J. Williams,\altaffilmark{1}
Kazimierz J. Borkowski,\altaffilmark{1}
Stephen P. Reynolds,\altaffilmark{1}
John C. Raymond,\altaffilmark{2}
Knox S. Long,\altaffilmark{3}
Jon Morse,\altaffilmark{4}
William P. Blair,\altaffilmark{5}
Parviz Ghavamian,\altaffilmark{5}
Ravi Sankrit,\altaffilmark{6}
Sean P. Hendrick,\altaffilmark{7}
R. Chris Smith,\altaffilmark{8}
Sean Points,\altaffilmark{8}
\& P. Frank Winkler\altaffilmark{9}
}

\altaffiltext{1}{Physics Dept., North Carolina State University., Raleigh, NC
    27695-8202; bjwilli2@ncsu.edu} 
\altaffiltext{2}{Harvard-Smithsonian Center for Astrophysics, 60
    Garden Street, Cambridge, MA 02138;}
\altaffiltext{3}{STScI, 3700
    San Martin Dr., Baltimore, MD 21218;}
\altaffiltext{4}{NASA Goddard Space Flight Center, Code 665, Greenbelt, 
    MD, 20771} 
\altaffiltext{5}{Dept. of Physics
    and Astronomy, Johns Hopkins University, 3400 N. Charles St., Baltimore, MD
    21218-2686;} 
\altaffiltext{6}{Space Sciences Laboratory,
    University of California, Berkeley, CA, 94720-7450;}
\altaffiltext{7}{Physics Dept., Millersville University, PO
    Box 1002, Millersville, PA 17551;} 
\altaffiltext{8}{CTIO, Cailla 603, La Serena, Chile;} 
\altaffiltext{9}{Dept. of Physics, Middlebury College,
    Middlebury, VT 05753;}

\begin{abstract}

We present near and mid-infrared observations of the pulsar-wind
nebula (PWN) B0540-69.3 and its associated supernova remnant made with
the {\it Spitzer Space Telescope}. We report detections of the PWN
with all four IRAC bands, the 24 $\mu$m band of MIPS, and the Infrared
Spectrograph (IRS).  We find no evidence of IR emission from the
X-ray/radio shell surrounding the PWN resulting from the forward shock
of the supernova blast wave. The flux of the PWN itself is dominated
by synchrotron emission at shorter (IRAC) wavelengths, with a warm
dust component longward of 20 $\mu$m. We show that this dust continuum
can be explained by a small amount ($\sim 1-3 \times 10^{-3} \msun$)
of dust at a temperature of $\sim 50-65$ K, heated by the shock wave
generated by the PWN being driven into the inner edge of the ejecta.
This is evidently dust synthesized in the supernova.  We also report
the detection of several lines in the spectrum of the PWN, and present
kinematic information about the PWN as determined from these
lines. Kinematics are consistent with previous optical studies of this
object. Line strengths are also broadly consistent with what one
expects from optical line strengths. We find that lines arise from
slow ($\sim 20$ km s$^{-1}$) shocks driven into oxygen-rich clumps in
the shell swept-up by an iron-nickel bubble, which have a density
contrast of $\sim 100-200$ relative to the bulk of the ejecta, and
that faster shocks ($\sim 250$ km s$^{-1}$) in the hydrogen envelope
are required to heat dust grains to observed temperatures. We infer
from estimates of heavy-element ejecta abundances that the progenitor
star was likely in the range of 20-25 $M_\odot$.

\end{abstract}


\section{Introduction}
\label{intro}

Many core-collapse supernovae (SNe) leave behind a neutron star as a
compact remnant.  Some of these neutron stars are active pulsars which
inflate a bubble of relativistic particles and magnetic fields
confined by the ejecta or interstellar medium (ISM), known as a
pulsar-wind nebula.  The combination of a shell supernova remnant
(SNR) and associated pulsar-wind nebula can allow the investigation of
various issues of importance in supernova and pulsar physics,
including pulsar kicks, ejecta structure and composition, and particle
acceleration at relativistic shocks.  Pulsar-wind nebulae serve as
calorimeters for pulsar spindown energy loss, and as test systems to
study the behavior of relativistic shocks where the pulsar wind is
thermalized. We know of few cases of a ``normal'' radio and X-ray
shell supernova remnant containing an active pulsar and synchrotron
nebula.  Probably the best known such ideal case is the Large
Magellanic Cloud remnant B0540-69.3 (or ``0540'' for short). 0540 is
also one of a highly exclusive group of ``oxygen-rich'' SNRs, a group
that includes Cas A, Puppis A, G292+1.8, 1E0102-72.3, and N132D.

Theoretical studies of PWNe have either concentrated on the gross
evolution, assuming a homogeneous nebula
\citep{rees74,pacini73,reynolds84} or the detailed spatial structure,
neglecting evolution \citep{kennel84}.  Since the advent of the
new generation of X-ray observatories, the study of PWNe has
accelerated, with the identification of many new objects and more
detailed information on known ones (see Gaensler \& Slane 2006 for a
recent review).  \cite{chevalier05} modeled PWNe for different
assumptions about the ejecta profiles into which they expand, to
relate properties of supernovae to those of the PWNe.
 
PWNe produce extremely broad-band spectral-energy distributions
(SEDs), well described in various frequency regimes with power laws.
Most PWNe are observed in radio and X-rays; only a few are detected
optically (here as in many other ways the Crab Nebula is an
exception), and almost nothing is known about infrared or ultraviolet
spectra.  Typical radio spectra are featureless, and are well described by
power-laws with spectral indices $\alpha < 0.3$ ($S_\nu \propto
\nu^{-\alpha}$), with X-ray indices steeper by 0.5 -- 1.3 (see data in
Chevalier 2005).  Since simple models of synchrotron losses predict a
steepening of exactly 0.5, they lack some essential physics, which may
be constrained if the complete spectrum is known.  Galactic PWNe are
all found close to the Galactic plane, where they suffer from
extinction in optical and UV and confusion in IR.  Filling in the SED
between radio and X-rays can best be done with a high-latitude object.
For this reason as for many others, 0540 is an interesting target.

0540 was first catalogued as a radio source of unknown nature, a minor
feature on a 408 MHz map of the 30 Dor region made with the Molonglo
telescope \citep{lemarne68}.  \cite{mathewson73} first classified it
as a supernova remnant on the basis of its steep radio spectrum, although
their optical survey did not detect it.  Early reports associated 0540 with 
the H$\alpha$ emission nebula N 158A 
\citep{henize56}, though that object is $3'$ from the centroid of the
early radio positions (which could be localized to better than
$10''$).  The absence of strong H$\alpha$ emission from 0540 further
demonstrates that the association with N 158A is erroneous.
Subsequent radio observations \citep{milne80} gave an improved
spectral index of $-0.44$, typical for a shell supernova remnant.  The
first indication of something unusual was the X-ray detection
\citep{long79} with the {\sl Einstein} Observatory, in which 0540 was
the third brightest X-ray remnant in the LMC.  The X-ray spectrum was
shown to be featureless by \cite{clark82} with the {\sl Einstein}
Solid-State Spectrometer.  The first optical detection was reported by
Mathewson et al.~(1980), motivated by pre-publication reports of the
observations of Clark et al~(1982).  Mathewson et al.~did not see
H$\alpha$ but instead a spectacular ring in [O III] of $8''$ diameter,
with a smaller ring in fainter [N II] emission ($4''$ diameter) and no
appreciable Balmer emission. In addition to classifying 0540 as an
``oxygen-rich'' SNR, \cite{mathewson80} also reported spectroscopic
observations indicating expansion speeds of order 1500 km s$^{-1}$.
The discovery of the 50 ms X-ray pulsar \citep{seward84} and optical
synchrotron nebula \citep{chanan84} added to the complexity and
interest of the system.  The optical emission was shown definitively
to be synchrotron by the discovery of polarization \citep{chanan90}.
The pulsar spindown timescale $P/2 {\dot P}$ is about 1660 yr
\citep{seward84}, somewhat longer than the kinematic age estimate
resulting from dividing the radius ($4'' = 1$ pc at our assumed
distance of 50 kpc) by the expansion speed of about 1500 km s$^{-1}$,
which yields a value of $\sim 700$ yr.  The pulsar spindown luminosity
is $1.5 \times 10^{38}$ erg s$^{-1}$.

\cite{reynolds85} modeled 0540 with the formalism of
\cite{reynolds84}, with the pulsar driving an accelerating synchrotron
nebula into the inner edge of expanding ejecta.  At that time, there
were no more than hints of extended structure that could be identified
with the outer blast wave.  \cite{reynolds85} found that the current
radio, optical, and X-ray observations could be explained without
requiring extreme values for the pulsar initial energy or other
parameters.  He deduced an initial pulsar period of about $30 \pm 8$
ms, that is, relatively slow, and concluded that the true age of 0540
was between 800 and 1100 yr, somewhat longer than the kinematic age
due to the pulsar-driven acceleration.

Up to this time, all observations were consistent with 0540 being a
standard Crab-like remnant (i.e., a nonthermal center-brightened radio
and X-ray nebula surrounding a pulsar), except for the hint of
larger-scale structure from radio images and from X-ray observations
\citep{seward84}.  Definitive information on the structure came from
higher-resolution radio observations with the Australia Telescope
\citep{manchester93} which showed a clear radio shell with diameter
about $65''$ surrounding a radio nebula with size (about $5''$ FWHM)
comparable to the bright X-ray nebula and [O III] ring.  The shell has
a radio spectral index $\alpha$ of about $-0.4$, while the central
nebula has $\alpha = -0.25$.  At this point it was clear that 0540 is
even more Crab-like than the Crab, as it possesses a clear outer blast
wave interacting with surrounding material, so that we could be sure
that the interior PWN is interacting with the inner SN ejecta as in
Reynolds \& Chevalier (1984).  X-ray emission from the blast wave was
confirmed with {\it Chandra} observations \citep{hwang01}; the
emission is brightest in the W and SW, like the radio.  Spectral fits
indicated abundances typical of the LMC, with a temperature of order 4
keV (for a Sedov blast wave model) and ionization timescale $\tau
\equiv n_e t = 3.7 \times 10^{10}$ cm$^{-3}$ s, though spectral
differences are apparent in different regions, and a hard component
may be called for.

The most thorough optical spectroscopic study to date was reported by
\cite{kirshner89}.  They confirmed the high velocities (FWZI $\sim
2800$ km s$^{-1}$), and reported weak H$\alpha$ emission.  The average
centroid of SNR lines (as opposed to narrower lines from a nearby H II
region) was shifted by $+370$ km s$^{-1}$.  No [Ne III] was reported ($<
1.5$ \% of [O III]); they concluded that this was a real abundance
deficit rather than a temperature or density effect.  A detailed study
by \cite{serafimovich04}, focusing on the optical nonthermal
continuum, revised the reddening and optical slope to give a power-law
index in the optical of $\alpha_o = -1.07.$ Recent observations by
\cite{morse06} report the discovery of faint [O III] emission
extending to a radius of $8''$, with a velocity of 1650 km s$^{-1}$.
They find the centroid of this velocity component to be the same as
that of the LMC, so that a large peculiar velocity of the system is
not required.

\cite{chevalier05} modeled 0540, along with several other PWNe, with
the goal of learning more about the SN explosion.  He obtained several
results for a simple dynamical model of a PWN expanding into ejecta of
various density profiles driven by a pulsar of given power. He
interpreted 0540 as the result of a SN Ib/c, an exploding Wolf-Rayet
star, with the prediction of a lack of significant emission from
hydrogen. However, recent optical observations by
\citet{serafimovich04} and \citet{morse06} have detected hydrogen. In
light of this, it is now believed \citep{chevalier06} that 0540 is the
result of a type IIP supernova.

The infrared observations of 0540, which was detected by the {\it
Infrared Space Observatory} (ISO) \citep{gallant99}, presented in this
paper promise to advance our understanding on several fronts. The
outline of our paper is as follows: In section 2, we describe the
observations and data reduction, and results are given in section
3. In section 4.1, we discuss a general picture of the PWN, and
sections 4.2 and 4.3 discuss in detail the line emission and dust
continuum emission, respectively. In section 4.4, we discuss the
origin of the O-rich clumps, whose existence we posit in section
4.2. Section 5 serves as a summary of our findings.

\section{Observations and Data Reduction}
\label{obs}

During Cycle 1 of {\it Spitzer} observations, we obtained pointed
observations of 0540 with the Infrared Array Camera (IRAC) and the
Multiband Imaging Photometer for Spitzer (MIPS) as part of a survey of
$\sim 40$ known supernova remnants in the Large and Small Magellanic
Clouds \citep{borkowski06,williams06}. Our IRAC observations (28
November 2004) consisted of a dither pattern of 5 pointings with a
frame time of 30 seconds for each frame. This pattern was used for all
4 IRAC channels. Our MIPS observations (7 March 2005) differed based
on the module used. At 24 $\mu$m, we mapped the region with 42
overlapping pointings of 10 seconds each. At 70 $\mu$m, we mapped the
remnant with 94 pointings of 10 seconds each. At 160 $\mu$m, we mapped
the region with 252 pointings of 3.15 seconds. Since 0540 was not
detected at 160 $\mu$m, we do not discuss 160 $\mu$m data here. Both
the Basic Calibrated Data (BCD) and Post-BCD products were processed
with version S14.4 of the PBCD pipeline. We then used the {\it Spitzer
Science Center} (SSC) contributed software package MOPEX to ``clean
up'' the images, although the improvements were minimal. MOPEX was
able to remove some of the streaks caused by bright stars in the IRAC
images of the region.

\begin{figure}
\plotone{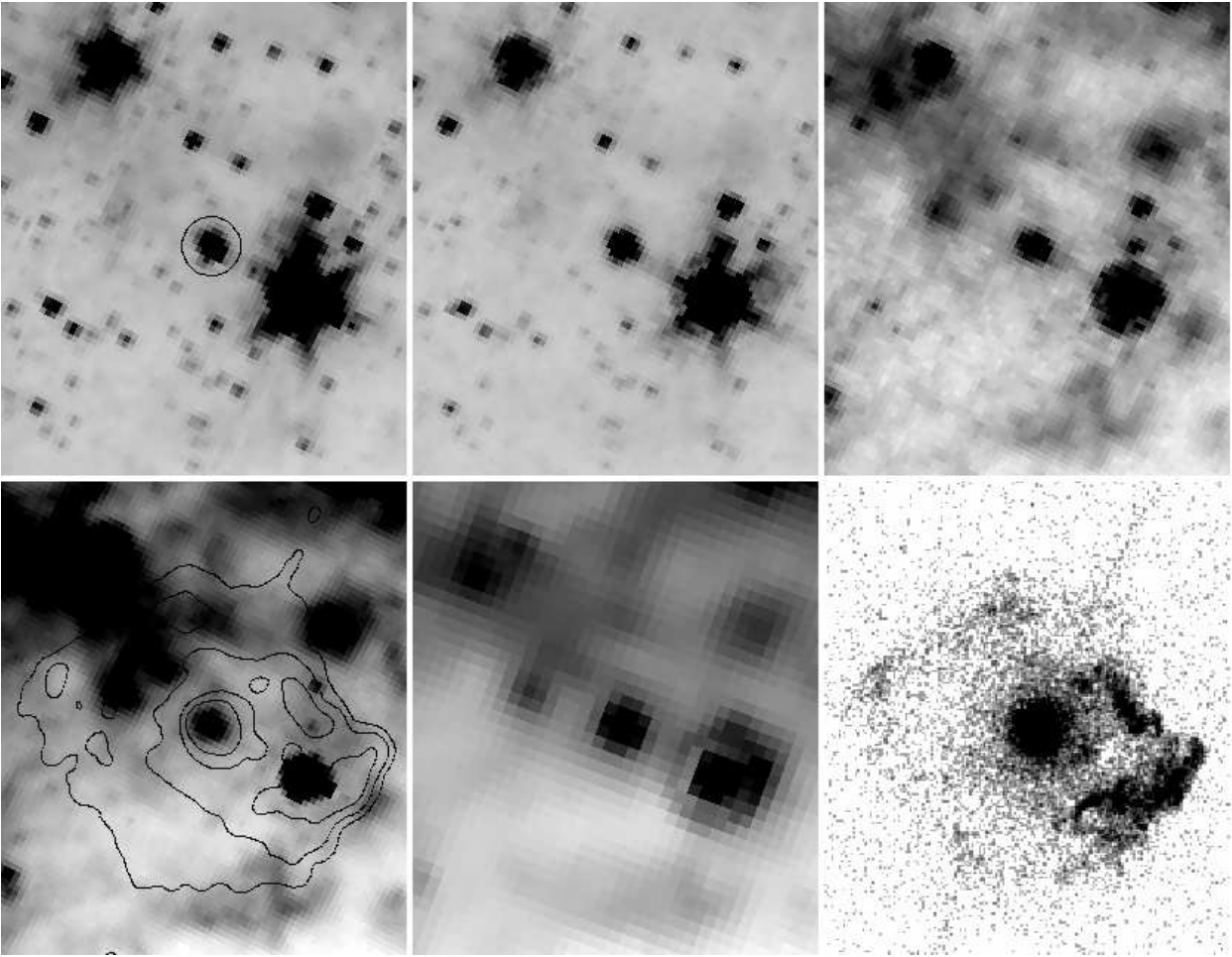}
\caption{Images of PWN 0540-69.3. Each image is approximately 100
arcseconds across. Left to Right, Top to Bottom: IRAC Chs. 1-4 (3.6,
4.5, 5.6, and 8.0 $\mu$m, respectively), MIPS 24 $\mu$m, {\it Chandra}
broadband X-ray image. The location of the PWN is marked with a circle
in the IRAC Ch.1 image, and X-ray contours are overlaid on the IRAC
Ch.4 image.}
\label{images}

\end{figure}

Our images of the source are shown in Figure~\ref{images}. With a
radius of $\sim 4''$, the PWN is resolved by {\it Spitzer}, and it
clearly stands out from the background in IRAC and MIPS 24 $\mu$m
bands. In IRAC ch. 3 \& 4 (5.8 \& 8.0 $\mu$m), as well as MIPS 24
$\mu$m, there is a hint of a shell around the nebula, at approximately
$30''$. We considered the possibility that this shell is related to the
SNR, perhaps the collisionally heated dust from the outer blast wave,
as we have observed in several other SNRs. However, the morphology of
the IR shell does not correspond with any features in the X-ray or
radio shell. Spectroscopy of the shell shows it to be virtually
identical to the surrounding background unrelated to the remnant, so we
are forced to conclude that its apparent relation to the SNR is
coincidental.

\begin{figure}
\plotone{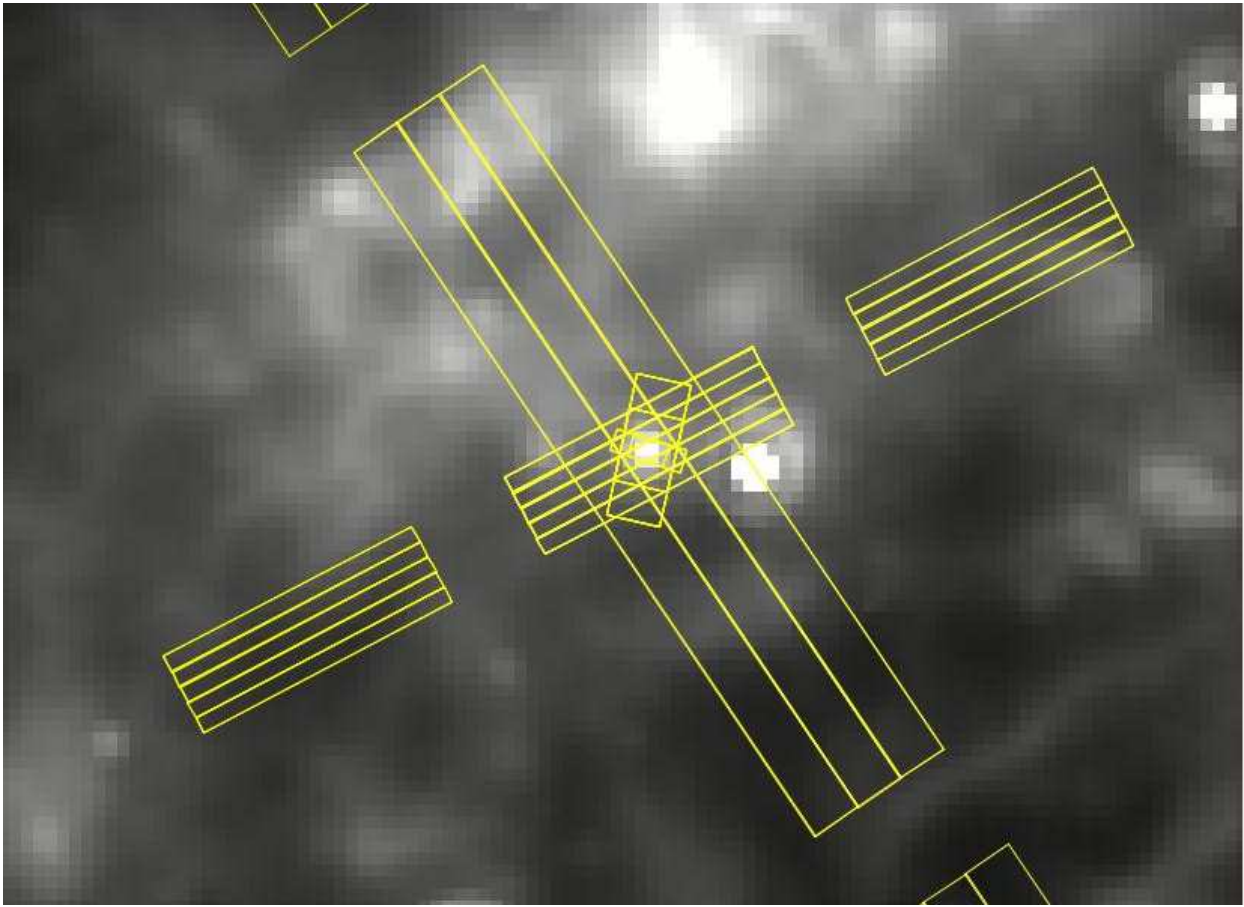}
\caption{Coverage of IRS slits overlaid on MIPS 24 $\mu$m image.}
\label{coverage}

\end{figure}

In Cycle 2, we obtained spectroscopic pointings for 0540 using all
four instruments of the Infrared Spectrograph (IRS). Our observations
were done between 8-10 July 2005. We used the spectral mapping mode
for the low-resolution modules, and staring mode for the
high-resolution echelle modules. Figure ~\ref{coverage} shows our
coverage of the PWN with IRS overlaid on our MIPS 24 $\mu$m image. For
the short-wavelength, low-resolution module (SL) we obtained 5
parallel pointings with each of the two orders, with a step direction
of $3.5"$ perpendicular to the dispersion direction of the slit. A
total of 480 seconds (2 cycles of 240 s) was obtained for each slit
position. For the long-wavelength, low-resolution (LL) module, we
obtained 3 parallel pointings for each order with a step direction of
$10.5"$ perpendicular to the dispersion direction. A single cycle of
120 seconds was used for LL. Since we are primarily interested in
determining the shape of the continuum from the low resolution
spectra, it was important to obtain spectra of the local background as
well as the source. Figure~\ref{images} illustrates the complex nature
of the local background and the difficulty of accurate background
subtraction. Because our source is only $\sim 8''$ in diameter, we
were able to extract background spectra from not only the parallel
slit pointings, but also from different parts of the slit containing
the source. We downloaded the Post-BCD data, pipeline version S15,
from the SSC. We used the Spitzer IRS Custom Extraction (SPICE)
software provided by the SSC to extract our spectra. Although the PWN
is slightly extended, it is close enough to a point source, especially
at longer wavelengths, to use the point source extraction mode in
SPICE.

In order to determine line profiles and strengths from the source
itself, we also obtained pointings with both the short-wavelength,
high-resolution (SH) and long-wavelength, high-resolution (LH) modules
in staring mode.  For these pointings, we centered the echelle
spectrographs only on the source, without a dedicated
background pointing. Since staring mode automatically provides 2 nod
positions for each pointing, we averaged the two to obtain a single
spectrum for SH and a single spectrum for LH. For SH, we used 3 ramp
cycles of 480 seconds each, for a total of 1440 s. The same total
integration time was obtained for LH, but was broken up into 6 cycles
of 240 s each.

\section{Results}
\label{results}

\subsection{Flux Extraction}

0540-69.3 is located close to the 30 Doradus region of the LMC, and
thus is in a region of high infrared background. For our IRAC and MIPS
images, we simply used an annular background region to subtract off
the background flux from the PWN. Because the nebula is only about $4''$
in radius, we used an on-source region of $\sim 6''$ radius to be sure
to capture all of the flux from the object, and a background annulus
between $6''$ and $10''$ radius. At 70 $\mu$m, we derive an upper limit to the
flux that could be contained in the region based purely on error
analysis of the pixels. Our results, with a 3$\sigma$ upper limit at
70 $\mu$m, are given in Table~\ref{fluxtable}.

\subsection{Spectral Extraction}

\begin{figure}
\plotone{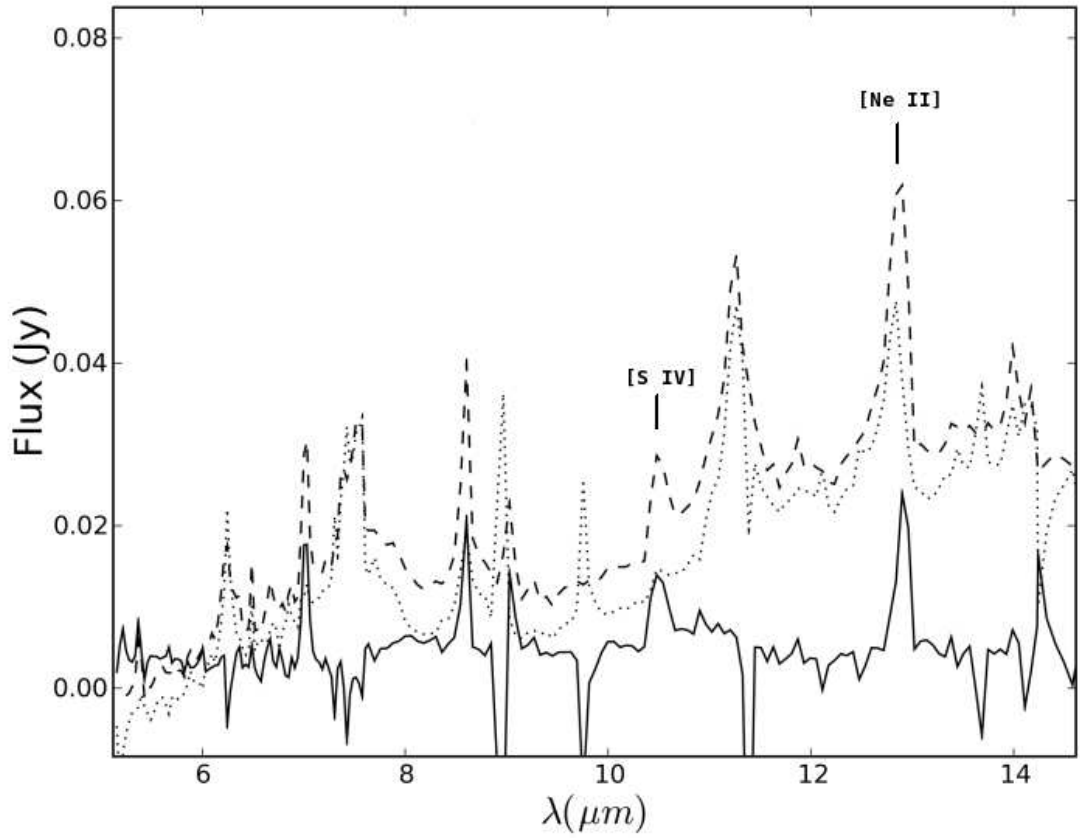}
\caption{The short-wavelength, low-resolution spectrum of the
PWN. Local background has been subtracted as described in the
text. Dashed line is source $+$ background; dotted line is background;
solid line is the spectrum of the source only.}
\label{sltotal}
\end{figure}

For each of the SL orders, our procedure was as follows. First, we
extracted spectra from three different, non-overlapping positions on
each of the 5 slits. The positions corresponded to the middle and the
ends of each slit. This gave us a total of 15 different spectra. Given
the spatially varying background in the vicinity of the PWN, we
elected to use only the background regions that were closest to the
source. Thus, we excluded the 4 ``corner" regions, leaving us with 11
total regions. We considered the middle 3 regions to be our
``on-source" region (since flux from the PWN was extended into all 3)
and added them together. We then used the remaining regions as
background. As a check of this method, we integrated the background
subtracted spectra over the appropriate wavelengths corresponding to
the 5.8 and 8.0 $\mu$m IRAC channels, factoring in the spectral
response curves. Within errors, we obtained the same flux here as we
did using aperture photometry on the IRAC images. In
Figure~\ref{sltotal}, we show the short-low spectrum of the PWN with
the both the original source spectrum and background spectrum
overlaid.

\begin{figure}
\plotone{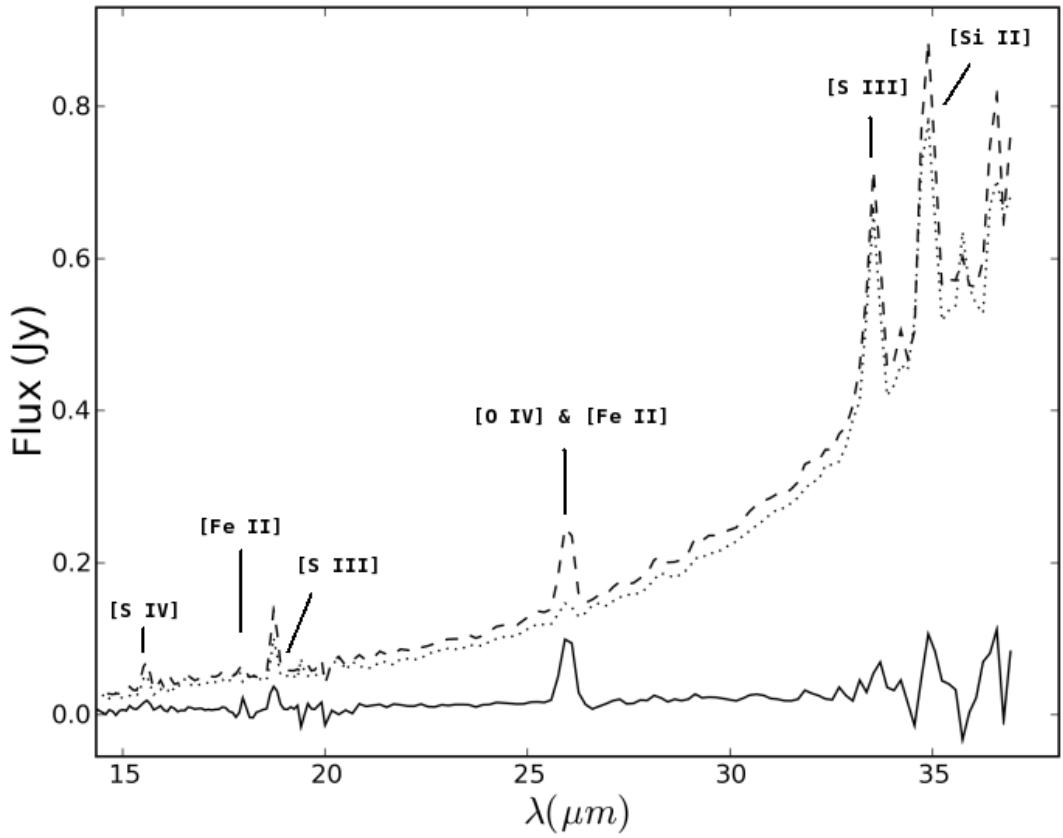}
\caption{The long-wavelength, low-resolution spectrum of the
PWN. Lines are the same as in Figure 3.}
\label{lltotal}
\end{figure}

For the LL slits, we followed a similar procedure. Because we only had
3 parallel slit positions, we had 9 total spectra extracted from
spatially different areas. Because the LL slit is wider than SL (about
$10.5''$), we only considered the middle region of the middle slit
to be the on-source region. In keeping with our policy of only using
the closest background regions, we again excluded the 4 corner
regions, and only used the 4 regions corresponding to the 2 middle
regions of the parallel slit pointings, and the 2 extractions from the
ends of the middle slit. We then averaged the 4 background spectra and
subtracted the result from the on source spectrum to get a background
subtracted spectrum. Again, as a check on this method, we integrated
the resulting spectrum over the appropriate wavelengths, and with the
appropriate spectral response curves, calculated a 24 $\mu$m flux
that could be compared with that derived from aperture photometry on
the MIPS image. Within errors, there was excellent agreement between
these two methods. Figure~\ref{lltotal} shows the long-low spectrum of
the PWN.

In order to examine the shape of the synchrotron continuum from the
low-resolution data, it was necessary to remove the lines from the
spectra, as well as artifacts produced by obviously bad
pixels. Although we detected the PWN at all wavelengths, the spatially
varying background made our background subtraction procedure somewhat
uncertain. Using different background subtraction regions does produce
different results for the final spectrum, mainly due to the steep
north-south gradient in the infrared background in the region of 0540
(see Figure~\ref{coverage}). We believe our approach of defining an
``annulus'' region and averaging the backgrounds around the source is
the best solution to this problem. However, it is not without
significant uncertainties. We tried several variations of different
background regions to see what the effects were. The largest
differences came in comparing the two same-slit background positions
with the two parallel slit background positions. We found variations
in the absolute flux level between these two choices to be on the
order of 40\%. Because we have no reason to favor one over the other,
we averaged them together with equal weights, thus creating our
annulus. Because of this, we have used caution in interpreting the
results of the extraction of the continuum. We also considered the
possibility that a large number of weak lines could be interpreted as
continuum. We reject this hypothesis for two reasons. First, there are
only a handful of IR lines predicted in this wavelength range, and
with the exception of [Ar III] at 8.99 $\mu$m, none of them even come
close to the detection limit based on our line models. Second, we have
high-resolution spectra of this region, and there is no evidence of
lines that would be unresolved in the low-resolution data.

\subsection{Line Fitting}

\begin{figure}
\plotone{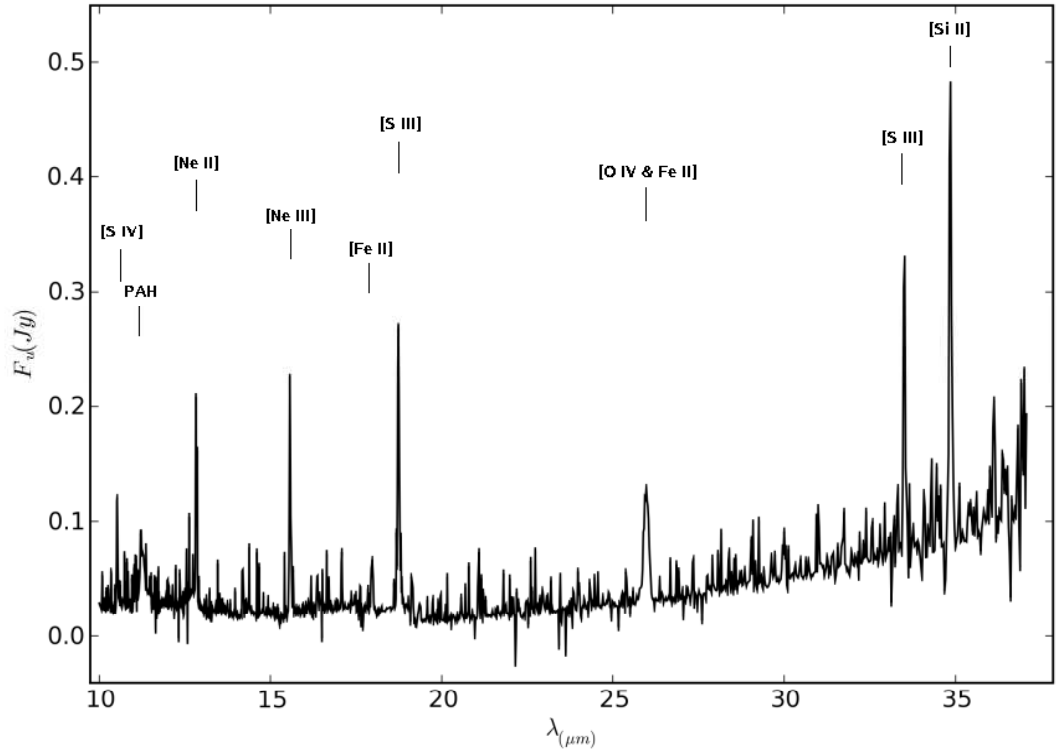}
\caption{The high-resolution spectrum of the PWN, with no background
subtraction. Measured lines are marked, along with a dust feature at
${\sim 11} \mu$m.}
\label{hightotal}
\end{figure}

We used SPICE to extract spectra from the high-resolution data as
well. In order to fit the lines, we used the open-source software
Peak-O-Mat, which runs on SciPy (Scientific Python) and is available
from http://lorentz.sourceforge.net/. Peak-O-Mat is an interactive
program that is designed to fit curves using a least-squares algorithm
to a user-specified function. Because our extraction region contains
not only the entire expanding shell of the PWN, but also the
foreground and background emission from the surrounding ISM, we
expected to see both broad and narrow components for most of the lines
detected, as has been seen in optical spectroscopy of the nebula. We
assumed Gaussian profiles for both the broad and narrow components,
and fit these on top of a linear background. We manually removed
artifacts that were clearly caused by bad pixels, as determined by
examining the 2-D dispersed image. We also clipped bad pixels from the
backgrounds in the vicinity of each line, in order to make the fitting
of the actual lines easier with a longer tail for the Gaussian. We did
not remove or alter any of the pixels that were contained in the line
itself, except in the case of the [Ne II] line at 12.8 $\mu$m. There
was an obvious bad pixel that was contaminating the line structure at
around 12.86 $\mu$m. In order to correct for this, we interpolated the
strength of that pixel based on the strengths of neighboring
wavelength pixels.  Line profiles and strengths are discussed in
section~\ref{disc}. The complete high-resolution spectrum of the PWN
is shown in Figure~\ref{hightotal}.

\begin{figure}
\plotone{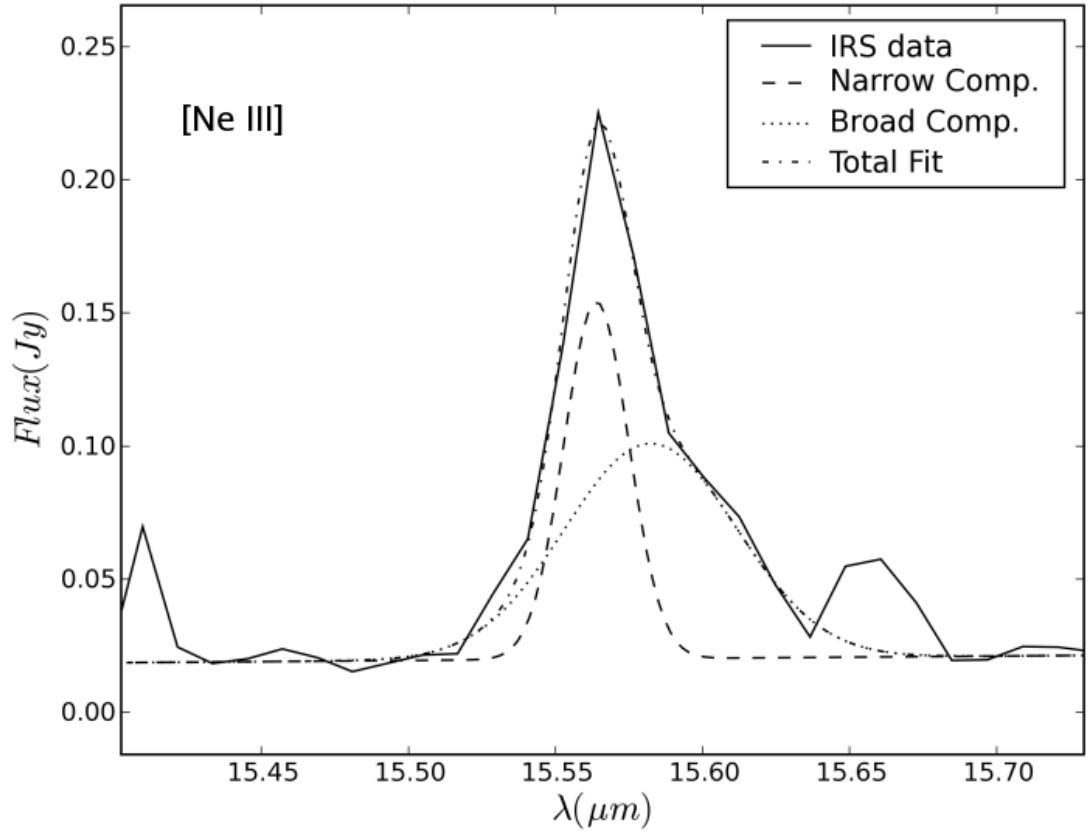}
\caption{An example of our two-component fit to the lines identified
in the high-resolution spectrum of the PWN. [Ne III] is clearly seen
to have two components. Noisy pixels were clipped out for the fitting,
but were left in this image to show their relative level of
contribution.}
\label{neIIIline}
\end{figure}

We find that nearly all of the lines in the spectrum have a
two-component nature, with a narrow component we attribute to the
surrounding H II region, and a broad component coming from the
PWN. Figure~\ref{neIIIline} shows an example of a two-component fit to
a line, in this case [Ne III], at 15.5 $\mu$m. The spectral resolution
of both SH and LH is $\lambda/\Delta\lambda$ $\sim 600$, which
corresponds to a minimum FWHM of 500 km s$^{-1}$. Since we do not expect the
narrow component widths to be wider than this, we fixed the narrow
component widths to this value. Furthermore, the LMC has an overall
recession velocity relative to the Sun of +270 km s$^{-1}$, so all narrow
components should be redshifted by this amount. However, when we fixed
the centroid of the Gaussian for the narrow component to this
velocity, the fits were unacceptably poor. According to the { \it
Spitzer Observer's Manual}, the wavelength calibration in IRS is 1/5
of a resolution element, which for the high-resolution module
corresponds to 0.003-0.011 $\mu$m, or 100 km s$^{-1}$. Since we found that
all the narrow components seem to be off by a comparable systematic
shift, we believe that the uncertainties in wavelength calibration are
responsible. Thus, we measured the shift for each narrow component and
averaged them to obtain a value to which we would fix each narrow
component. We considered SH and LH separately, and calculated that
each narrow component was redshifted on average (relative to its rest
wavelength) 171 km s$^{-1}$ for SH and 230.5 km s$^{-1}$ for LH. Fixing the
centroids of the narrow components to these values returned much more
acceptable fits.

After we used Peak-O-Mat to determine the best values for the
parameters of either one or two Gaussians, we then used our own
least-squares algorithm to obtain errors. The errors listed on the
parameters in Table~\ref{linetable} are 90\% confidence limits,
corresponding to a rise in $\chi^{2}$ of 2.706 from its minimum
value. This procedure was repeated for each parameter
separately. Errors on line fluxes were obtained through the standard
error propagation formula.

\section{Discussion}
\label{disc}

\subsection{General Picture}

We aim at a self-consistent, semi-quantitative picture of the PWN that
accounts for the presence of lines (optical and IR), the extent of the
synchrotron nebula, and the source of the [O III] emission at $8''$
radius. We find it useful to first point out some contrasts between
0540 and the most widely-known object of its class, the Crab
Nebula. Although 0540 has been referred to as ``The Crab's Twin,'' the
two differ in some important ways. The most obvious difference is the
lack of an outer shell in the Crab, while 0540's $30''$ shell has been
seen in both radio and X-ray observations. For the purposes of this
paper, however, the important differences lie in the PWN. In the Crab,
the size of the nebula decreases with increasing frequency, so that
the radio nebula is larger than the optical, which is larger than the
X-ray, etc. In 0540, the synchrotron nebula is approximately identical
in extent throughout all wavelengths, around $5''$. The other
fundamental difference is the presence in 0540 of emission located
beyond the synchrotron nebula (the [O III] halo). There is nothing
like this seen in the Crab, where the radio synchrotron emission
extends to the outer boundary of anything known to be associated with
the nebula.

In modeling the Crab Nebula, \cite{sankrit97} considered two models,
one a pure shock model and the other a pure photoionization model to
explain the optical emission. They concluded that shocks from an
expanding shell were more likely. In the case of 0540, however, a pure
shock model cannot reproduce the [O III] extended emission. We
therefore propose an extension to their models that incorporates both
a global shock {\it and} photoionization. The specifics of our model
will be described more fully in the sections below, but our general
picture of the nebula is as follows.  It is based on the dynamical
picture of Chevalier (2005; C05).

Approximately a millenium ago, a star exploded via the core-collapse
mechanism, leaving behind a pulsar, and sending a shock wave out into
the interstellar medium. The outer boundary of this forward shock is
now about 8 pc (angular distance of about $30''$) from the pulsar, and
the reverse shock into the ejecta is somewhere between $10'' - 30''$,
having not yet reached back to the inner ejecta. The pulsar has since
formed a pulsar-wind nebula, which itself is driving a shock into the
inner edge of the surrounding ejecta, which are in free expansion. The
shock wave heats the inner ejecta and sweeps them into a thin
shell. Since the shell of material is being continuously injected with
energy from the pulsar, it is accelerating and overtaking less dense
material as it expands. The shock speed relative to upstream material,
however, reaches a maximum and then begins to drop since the
free-expansion speed of the ejecta material is also higher at larger
radii. There is no reason, however, to expect the ejecta to be
completely homogeneous. The $^{56}$Ni synthesized in the explosion
will have heated the central ejecta by radioactive decay, causing them
to expand in an ``iron-nickel bubble'' \citep{li93}, and compressing
intermediate-mass ejecta into a denser surrounding shell.

We propose that the PWN shock has reached a radius of about 1.2 pc
from the pulsar, which corresponds to a size of $\sim 5''$, the size
of the nebula as determined by X-ray observations. The layer of
shocked ejecta is geometrically thin, bounded on the inside by a
contact discontinuity separating it from the the PWN proper, which is
the shocked pulsar wind. The shock has already encountered and
propagated through the low-density iron-nickel bubble and its
surrounding shell. That shell is likely to be highly clumpy
\citep{basko94}; shocks driven into the clumps of heavy-element ejecta
will be slow. Finally, at a sub-arcsecond radius we expect the
inward-facing pulsar wind shock where the relativistic pulsar wind is
thermalized. Interior to the shock driven into the ejecta, emission in
optical/IR is both thermal and non-thermal, with the dominant
component being synchrotron continuum emission from the relativistic
electrons.  However, multiple emission lines are clearly detected from
dense clumps and filaments of thermal gas.  In addition to this, we
identify a rising continuum in the mid-infrared above the synchrotron
continuum that we interpret as a small amount of warm dust,
collisionally heated by electrons heated by the shock. Most lines seen
in optical and infrared then come from dense clumps of ejecta, where
the shock wave has slowed significantly and become highly radiative.

What remains is to explain the faint [O III] emission seen at
$8''$. We propose that this is material that is still in free
expansion, i.e. unshocked, that has been photoionized by ultraviolet
photons emitted from the shockwave. The source of photoionization is
two-fold; ultraviolet photons from the synchrotron nebula and those
produced in fast radiative shocks both contribute appreciable amounts
of ionizing radiation. We show below that to within a factor of 2,
there are enough ionizing photons produced to account for the [O III]
halo at $8''$.

\begin{figure}
\plotone{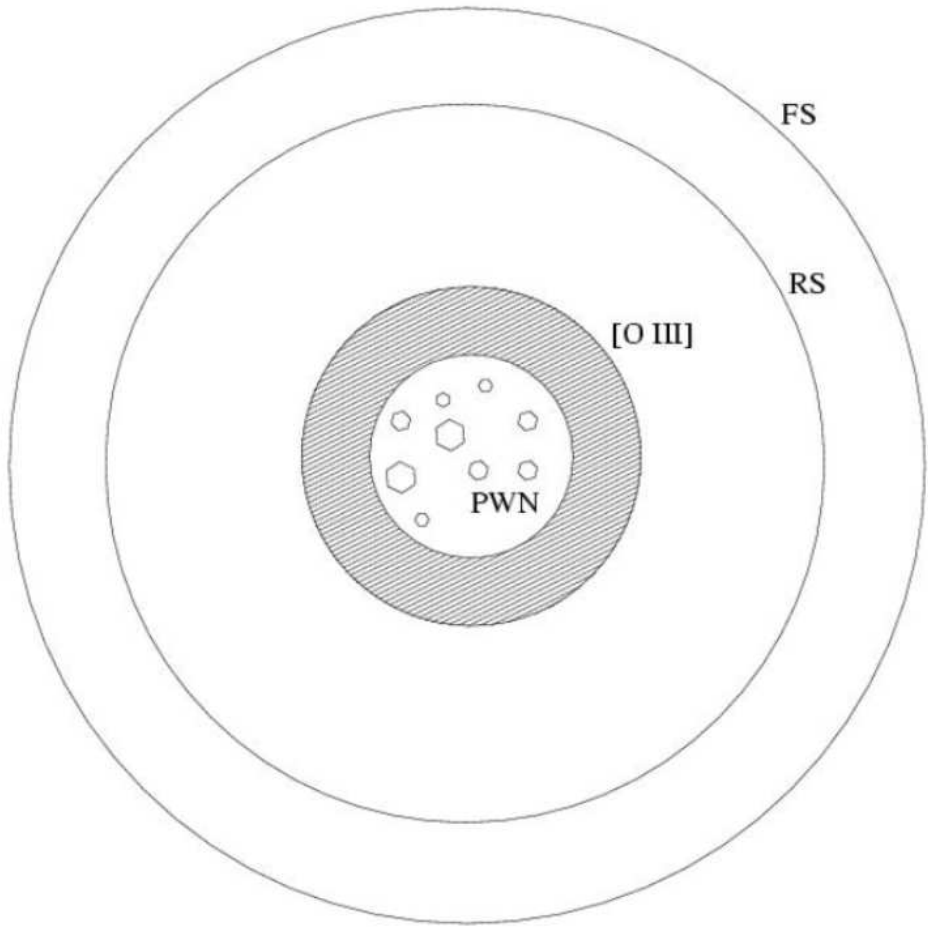}
\caption{A cartoon sketch of our general picture discussed in section
  4.1. Not to scale. FS refers to the forward shock from the SN blast
  wave, at a radius of $\sim 30''$. RS refers to the reverse shock,
  which has not yet been observed, and is at an unknown position
  between 10 and 30$''$ from the pulsar. [O III] refers to the extent
  of the halo of material that has been photoionized, and is seen in
  optical images to extend to 8$''$. PWN refers to the edge of the
  shock driven by the pulsar wind, and is located at a radius of
  5$''$. Interior to this shock, ejecta material has fragmented into
  clumps. The PWN as a whole is observed to have a redshifted velocity
  as reported in previous optical observations, possibly resulting
  from a pulsar kick. This is also the region where relativistic
  particles from the pulsar create observed synchrotron emission; see
  discussion in text.}
\label{cartoon}
\end{figure}

We have included, in Figure~\ref{cartoon}, a cartoon sketch of this
picture, which will be further discussed in the following sections. A
factor of a few is all we expect to be able to accomplish in modeling
the nebula, due to the large uncertainties involved.  These
uncertainties include, but are not limited to; nature of the
progenitor star (which affects the post-explosion density distribution
of the ejecta), heavy element abundances in the ejecta, degree of
clumping of the ejecta, etc. We have endeavored in the following
sections to point out places where uncertainties arise, and where
possible, to assign quantitative values to them.

\subsubsection{PWN Model}

C05 discusses a model, based on a thin-shell approximation, for a
pulsar wind nebula interacting with an inner supernova ejecta density
profile. We have used this model along with our observations to
determine various quantities about 0540, including how much hot gas
should be present. Observable quantities for the pulsar include period
($P$), period derivative ($\dot P$), luminosity ($\dot E$) and for the
nebula, size ($R$) and shell velocity ($V_{sh}$). While the quantities
for the pulsar are fairly well established by previous observations,
those for the PWN are much more uncertain. Previous optical studies of
the remnant \citep{mathewson80, kirshner89}, as well as radio
observations \citep{dickel02} interpreted the PWN as a bubble of
radius $\sim 4''$, and the optical observations gave expansion
velocities less than 1400 km s$^{-1}$. Based on {\it Chandra}
observations, \citet{petre07} concluded the nebula was slightly
larger, with a radius of $\sim 5''$. We shall adopt $5''$ (1.2 pc) as
an estimate of the location of the ejecta shock.

However, \citet{morse06} detected faint [O III] emission in images
extending out to a radius of $\sim 8''$. Based on similarities to the
Crab Nebula, they interpreted this [O III] halo as being the outer
edge of the shock from the pulsar wind overtaking the slower moving
ejecta. Here we present an alternative interpretation of this [O III]
halo emission. We propose that it is undecelerated ejecta that have
been photoionized, rather than shock-ionized. The FWZI of the [O III]
emission from Morse et al. was 3300 km s$^{-1}$, which, given the
extent of 1.8 pc and our interpretation of this as undecelerated
ejecta, provides the remnant age of 1140 years. While this is somewhat
longer than the favored model of \citet{reynolds85}, it is at least
reasonable given other age estimates made by previous studies of the
object. Photoionization calculations are discussed in Appendix A.

As a first attempt to model the observations, we considered the case
of a spherically symmetric shock wave driven by the energy input from
the pulsar expanding into a medium with density profile described by
$\rho_{SN} = At^{-3}(r/t)^{-m}$. We considered different values of the
parameter $m$, as dynamical mixing between the ejecta and surrounding
medium would produce a complicated density structure. The swept-up
mass does not exceed $1 \msun$ in this model. Although this model did
a reasonably good job at producing shock speeds high enough to account
for the necessary dust grain heating rate, a spherically symmetric
model does not adequately reproduce line radiation observed in both
optical and IR. A slow shock into dense material is required to
explain these lines, and the spherical model cannot account for this,
since presence of lines requires a departure from the overall
homogeneous density profile. We present the spherically symmetric
calculations in Appendix B. A more robust model is required to explain
both the slow shocks required for lines and the faster shocks required
for dust emission. We will return to this picture at the end of the
following section, but we must first describe our line observations in
detail.

\subsection{Lines}
\label{lines}

Eight emission lines are detected in the Spitzer spectrum.  They
provide constraints on the density and temperature of the emitting
gas, and perhaps more importantly on the elemental abundances.  They
complement the optical spectra published by \citet{kirshner89} (K89),
\citet{morse06} (M06) and \citet{serafimovich05} (S05). We first
summarize the implications of the optical spectra, then consider shock
wave models for the combined optical and IR emission.

Several temperature estimates are available from the optical spectra.
The [O III] line ratio I(4363)/I(5007) gives temperatures of about
24,000 K according to S05 or 34,000 K (K89).  According to the CHIANTI
database \citep{landi06}, the ratio given by K89 corresponds to 50,000
K, while that given in M06 implies 24,000 K.  K89 also find
temperatures $>$30,000 K from the [O II] I(7325)/I(3727) ratio and
$<$10,000 K from [S II] I(4072)/I(6723).  The [S II] ratio of M06
implies $T$ = 14,000 K.  Assuming a temperature of 10,000 K, S05 find
a density of 1400 -- 4300 $\rm cm^{-3}$, and at 14,000 K the range
would be 1700 -- 5000 $\rm cm^{-3}$.  The differences among the
various temperature estimates may result partly from different
reddening corrections and different slit positions, but it is clear
that the [O III] lines are formed in hotter gas than the [S II] lines.
The Spitzer data include only one pair of lines from a single ion, [Fe
II] I(17.9$\mu$)/I(26.0$\mu$), which is constrained to be larger than
1.13.  Again using CHIANTI, this requires a density above about 5000
$\rm cm^{-3}$ and a temperature above 4000 K.  However, the ratio
depends upon the deblending of the [Fe II] and [O IV] lines at
26$\mu$m, and the uncertainty may be larger than the formal value.
The density contrast between the optically emitting material and the
mean post-shock density from the global model indicates that as in the
Crab Nebula, optically emitting material is concentrated in dense
knots and/or filaments.

The next step in interpreting the spectra and constructing models is
to estimate the relative importance of photoionization and shock
heating.  In the Crab nebula, photoionization dominates, though shocks
are important for the UV lines produced at higher temperatures and for
compressing the gas to increase the optical emissivity
\citep{sankrit97}.  In the oxygen-rich SNRs, such as N132D and
1E0102-7219, shock heating dominates \citep{blair00}.  0540
shows both synchrotron emission reminiscent of the Crab and extreme
heavy element enhancement.  A pure photoionization model with strongly
enhanced abundances and the observed density gives too low a
temperature to account for the [O II], [O III] and [S II] line ratios,
while shock models cool so rapidly that they produce little [O I] or
[S II] unless they produce no [O III] at all. Therefore, it seems
likely that a model of a shock including the PWN ionizing radiation is
needed.

We have computed models with the shock model described in
\citet{blair00} illuminated by the power law continuum described by
\citet{serafimovich04}.  Briefly, the code is similar to that of
\citet{raymond79} and \citet{cox85}, but it has been modified to
describe SNR ejecta with little or no hydrogen.  The most important
difference is that the cooling rate is enormous, so that the electron
temperature is well below the ion temperature in the hotter parts of
the flow.  The model is similar to those of \citet{itoh81} and
\citet{sutherland95}.  Unlike those models, we do not include the
photoionization precursor of the shock, because the ionizing emission
from the shock is considerably weaker than the ambient synchrotron
radiation.  In comparison with the spectra of Cas A, N123D and
1E0102-7219, shock models have the problems that no single shock model
produces the observed range of ionization states, and that they tend
to predict too much emission in the O I 7774 \AA\/ recombination line
unless the cooling region is somewhat arbitrarily truncated (Itoh
1988).  However, they do predict reasonable relative intensities from
the UV to the near IR for O III and O II. Below, we attribute the
truncation to mixing with hotter, lower density gas.

We assume a 20 $\rm km~s^{-1}$ shock with a pre-shock density of 30
$\rm cm^{-3}$, which produces a density of around 5000 $\rm cm^{-3}$
where the [S II] lines are formed.  The elemental abundances are O:
Ne: Mg: Si: S: Ar: Ca: Fe = 1: 0.2: 0.1: 0.1: 0.1: 0.1: 0.1: 0.1 by
number.  H, He and N are not included in the model, because it seems
likely that the lines from these elements arise in some other gas,
like either the quasi-stationary flocculi or the outer shell of ejecta
seen as very fast knots in Cas A \citep{kirshner77,fesen01}.  The
normalization of the power law flux assumes that the shocked gas is 1
pc from the center of the PWN.

The list of caveats is long.  There is undoubtedly a range of shock
speeds and pre-shock densities.  The shocked gas is unlikely to be a
uniform mixture of the various elements, and large variations in the
composition among different clumps, as observed in Cas A, are likely.
There may well be a significant contribution from unshocked
photoionized gas for some lines \citep[e.g.,][]{blair89}, as we shall
argue below for the [O III] halo.  The shock models are plane
parallel, with the power law illumination incident from the PWN,
while the X-rays are more likely to illuminate the shocked gas from
behind.  The models terminate somewhat arbitrarily at 250 K because of
numerical limitations.  This will affect the IR lines and the O I
recombination line at 7774 \AA .  Also, as a compromise between energy
resolution and energy range of the ionizing radiation, the power law
only extends to 2 keV.  This means that the inner shell ionization and
Auger ionization of S and Fe is not included.  Finally, the atomic
data in the code are somewhat out of date and need to be updated.
Nevertheless, the code gives a reasonable idea of the relative line
intensities.

To compare this model with our observed IR spectra, we must place the
Spitzer spectrum on the same scale as the optical spectra.  We
normalize the IR lines to [O III] 5007 = 100 by dividing the Spitzer
intensities by 4 times the [O III] 5007 intensity given by M06.  The
factor of 4 is meant to account for the fact that the $2''$ slit used
by M06 covers only about 1/4 of the remnant.  This is obviously not a
very accurate correction, but it is probably good to a factor of 2.
Since the relative fluxes of many optical lines differ by a factor of
2 between M06 and K89, this is unfortunately the best we can do until
an optical spectrum of the entire remnant becomes available.

The result is shown in Table 3.  Overall, the agreement is
astonishingly good for such a simple model.  Several of the low
ionization lines, [O I], [Ne II] and [Si II] are underpredicted,
though the [Si II] line could be increased simply by increasing the
silicon abundance.  The ratio of the [S IV] to [S III] IR lines is too
low, but inclusion of the harder part of the power law spectrum would
improve that.  Inclusion of the harder X-rays would also increase the
intensity of the [Fe VII] line, though a lower pre-shock density or a
higher shock speed would have the same effect. The oxygen column
density of the model is only about $10^{14}~\rm cm^{-2}$, and the
thickness of the emitting region is only $6 \times 10^{11}$ cm.  If
the thickness were large enough to allow the remaining O$^+$ to
recombine, the predicted O I recombination line, which is comfortably
lower than the weakest detected lines, would increase to about 4 times
the apparent detection limit of K89.  The agreement would improve if
the argon abundance were cut in half, but otherwise the abundances
appear to match the observations.

The shock model shown in Table 2 produces $1.1 \times 10^{-14} ~\rm
erg~cm^{-2}~s^{-1}$ in the [Ne III] 15.5 micron emission line, so the
flux shown in Table 1 would come from a region with surface area of $2
\times 10^{38}$ cm$^{2}$. This area is roughly equal to the area of a
sphere of $5''$ radius but the emission could come from many smaller
volumes with a total filling fraction of a few percent. Heavy-element
ejecta enter these slow shocks at a rate of $\sim 0.01 \msun$
yr$^{-1}$.

We conclude that, as in the Crab \citep{sankrit97}, the observations
can be explained by shocks that heat and compress the gas in the
radiation field of the PWN.  The shock heating seems needed to reach
the high temperatures seen in some line ratios and to provide the high
densities observed, while the photoionization heating strengthens the
low and moderate ionization lines.  Oxygen is about ten times as
abundant as the other elements.  The shock speed and pre-shock density
are not very well constrained, but a shock as fast as 80 $\rm
km~s^{-1}$ requires a low pre-shock density to match the observed
density, and that in turn implies a very high pre-shock ionization
state, overly strong [O IV] and overly weak [O II] emission.

In the last 100 years alone, about 1 $\msun$ of heavy-element ejecta
have been shocked, more than the total mass of the swept-up ejecta in
the global model described in Appendix B. This casts some doubt on the
validity of the global, spherically symmetric model, where density
within freely expanding ejecta was assumed to be a smooth
power-law. It is possible that the innermost ejecta have been swept-up
by an iron-nickel bubble, as inferred for SN 1987A by \cite{li93} and
modeled by \cite{basko94} and \cite{wang05} (see also brief
discussion in C05). We explored the possibility that the global shock
could be contained within the shell swept-up by the iron-nickel
bubble. In this one-dimensional picture, the shock passed through the
inner, low-density region in $\sim 50$ years, and has since been
contained within the high-density ($n \sim 30$ cm$^{-3}$) bubble
wall. We varied parameters of the model until the shock speed in the
bubble was approximately 20 km s$^{-1}$, as required by line
models. We find, however, that the mass flux of material entering the
shock throughout the remnant's entire lifetime has been unreasonably
high for this model, approximately $0.01 (t/1140$ yr$)^{-1/2} \msun$
yr$^{-1}$. In addition, a 20 km s$^{-1}$ shock, even at such density,
would not adequately heat dust grains to temperatures observed. Dust
heating is discussed in more detail in section~\ref{grains}.

\begin{figure}
\plotone{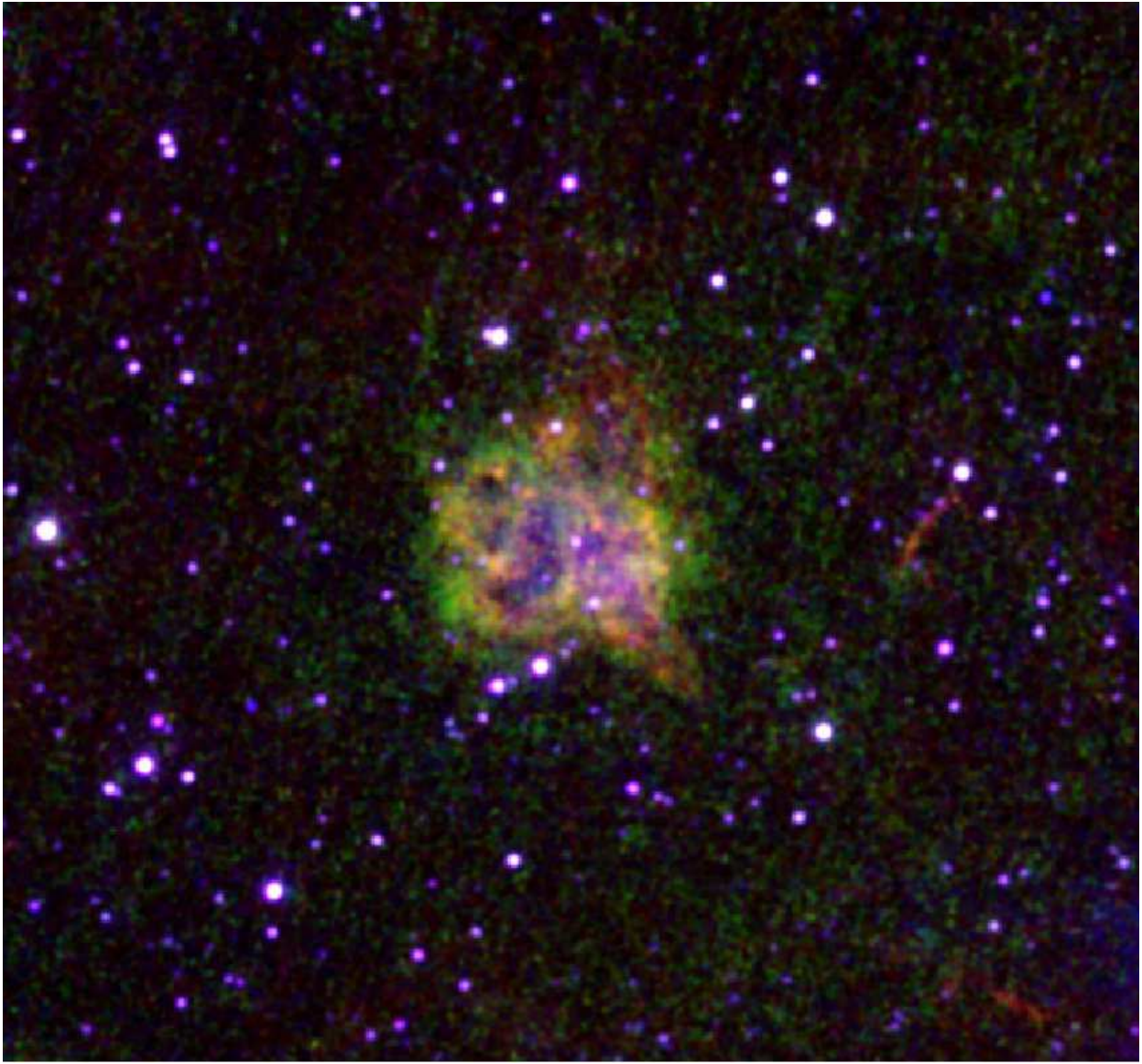}
\caption{{\it Hubble Space Telescope} WFPC2 image of PWN 0540-69.3, from
\cite{morse06}. Colors are: Blue - F791W continuum; Green - F502N [O
III]; Red - F673N [S II]}
\label{hstimage}
\end{figure}

We are forced to consider inhomogeneous ejecta with a fast global
shock to heat dust to observed temperatures, and slower shocks
producing observed line emission. We propose the following picture:
The shock swept through the low-density iron-nickel bubble interior
early in the life of the SNR. It then encountered the dense, clumpy
shell of the bubble, slowing down and further fragmenting the shell
into dense clumps, which are still being overrun by slow shocks,
currently 20 km s$^{-1}$. The global shock has now exited the
iron-nickel bubble shell, and is propagating through the ejecta with
relatively low ambient density. The speed of this shock is not well
known, but 250 km s$^{-1}$ would be sufficient to heat the dust to the
observed temperature of around 50 K (see below). Assuming pressure
equilibrium between the dense clumps and the ambient ejecta, we derive
a density contrast, given the difference in shock velocities, of $\sim
150$. Support for this model can be inferred from {\it HST} images of
the nebula, as seen in Figure~\ref{hstimage}, which shows [O III]
filaments in the interior, not just in a shell. [O III] line profiles
(M06) also do not match the shape that would be expected from a
spherically symmetric expanding shell, i.e. a flat top. The slow
shocks driven into the dense clumps are in rough pressure equilibrium
with the fast shock driven into the less dense ejecta.

\subsubsection{Progenitor Mass}
\label{progenitor}

We have compared the abundances of heavy elements listed in
section~\ref{lines} with the predicted abundances of \cite{woosley95},
who consider abundance yields from core-collapse SNe ranging in mass
from 11-40 $M_\odot$ and metallicities between zero and solar. We
consider models with both solar and 0.1 solar metallicity, as this
range is most likely to reflect a massive star in the LMC. Our
abundances listed are somewhat uncertain, and result from fits to
optical and infrared line strengths. We considered the ratios of O to
Ne, Mg, Si, and Fe. The data do not single out a particular model from
\cite{woosley95}, but ratios of heavy elements to oxygen do favor a
low-to-medium-mass progenitor. High-mass progenitors ($\gapprox$ 30
$M_\odot$) are less favored, since they produce larger amounts of
oxygen relative to other elements. This interpretation is consistent
with that of \cite{chevalier06}, who favored a type IIP explosion for
this object based on observations of hydrogen in the spectrum. This is
also consistent with the idea that type IIP SNe should result from the
explosion of a single star of 8-25 $M_\odot$ \citep{woosley02}.

It is possible to quantify these results even further. If we assume a
constant heavy-element mass flux through the radiative shocks for
$10^{3}$ yr of 0.01 $M_\odot$ yr$^{-1}$ with our abundances listed
above, we get a total ejected mass in oxygen of $\sim 3.5 M_\odot$,
though this number should only be considered accurate to a factor of a
few, and is likely an upper limit. When compared with predictions from
models, this value favors stars in the range of 20-25
$M_\odot$. \cite{maeder92} gives slightly different abundance yields
for SNe, with lower overall oxygen abundances produced. In his model,
high-mass stars ($\gapprox 25 M_\odot$) actually produce less oxygen
than their lower-mass counterparts, due to mass-loss of outer layers
and inability to synthesize O from He and C. However, these massive
stars would be Wolf-Rayet stars, and can be ruled out based on the
detection of hydrogen in optical spectra.

\subsection{Dust}

\begin{figure}
\plotone{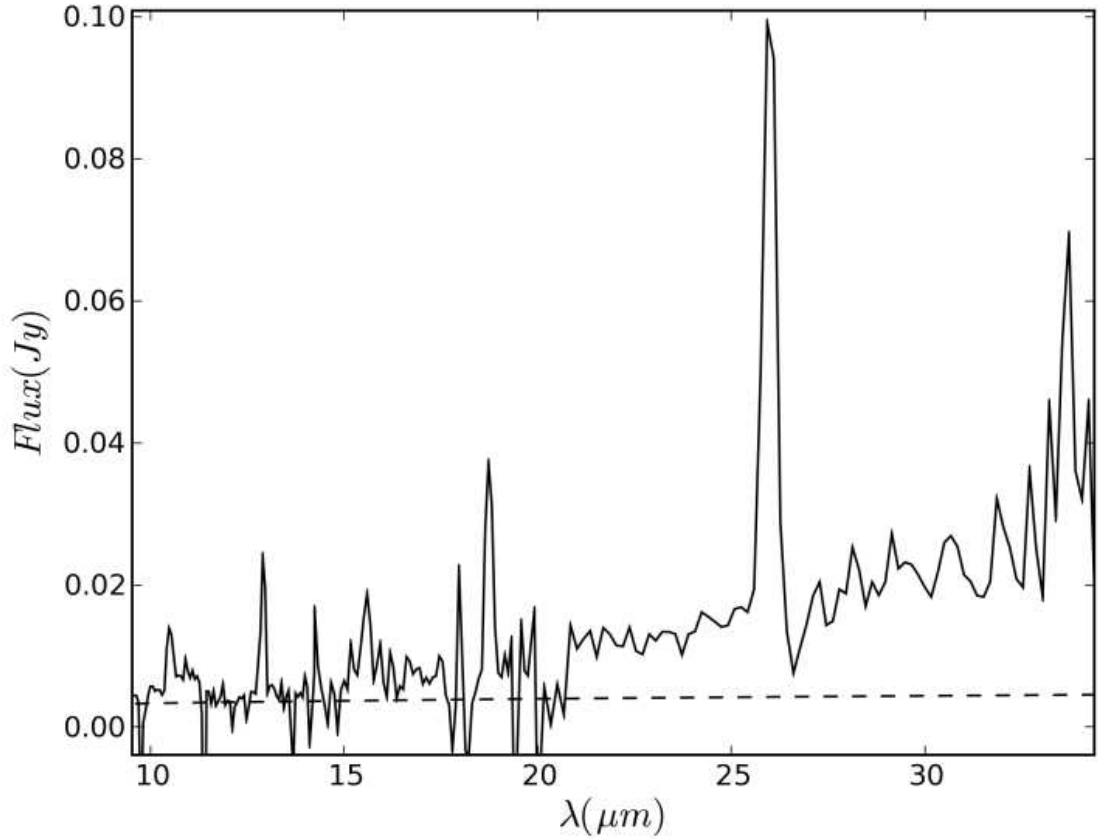}
\caption{The background-subtracted low-resolution spectrum of the PWN
is plotted as the solid line, with the radio synchrotron component shown as
a dashed line. A clear rising excess can be seen longward of 20 $\mu$m.}
\label{ll_sync}
\end{figure}

One of the more obvious features of the continuum in 0540 as seen in
Figure~\ref{ll_sync} is the excess of emission above the extrapolated
radio synchrotron spectrum at longer wavelengths. A similar excess has
been observed in the Crab \citep{temim06}, and has been attributed to
warm dust. We have inferred the temperature and the amount of dust
present, and have examined several possible mechanisms for grain
heating.

In order to fit the long-wavelength excess above the continuum, it was
necessary to remove contributions from emission lines and the
underlying synchrotron continuum. The flux contributed by the lines is
negligible, but their presence makes fitting of a model dust spectrum
more difficult. We have thus clipped obvious emission lines and bad
pixels out of the spectrum for this analysis.

\subsubsection{Synchrotron Component}

\begin{figure}
\plotone{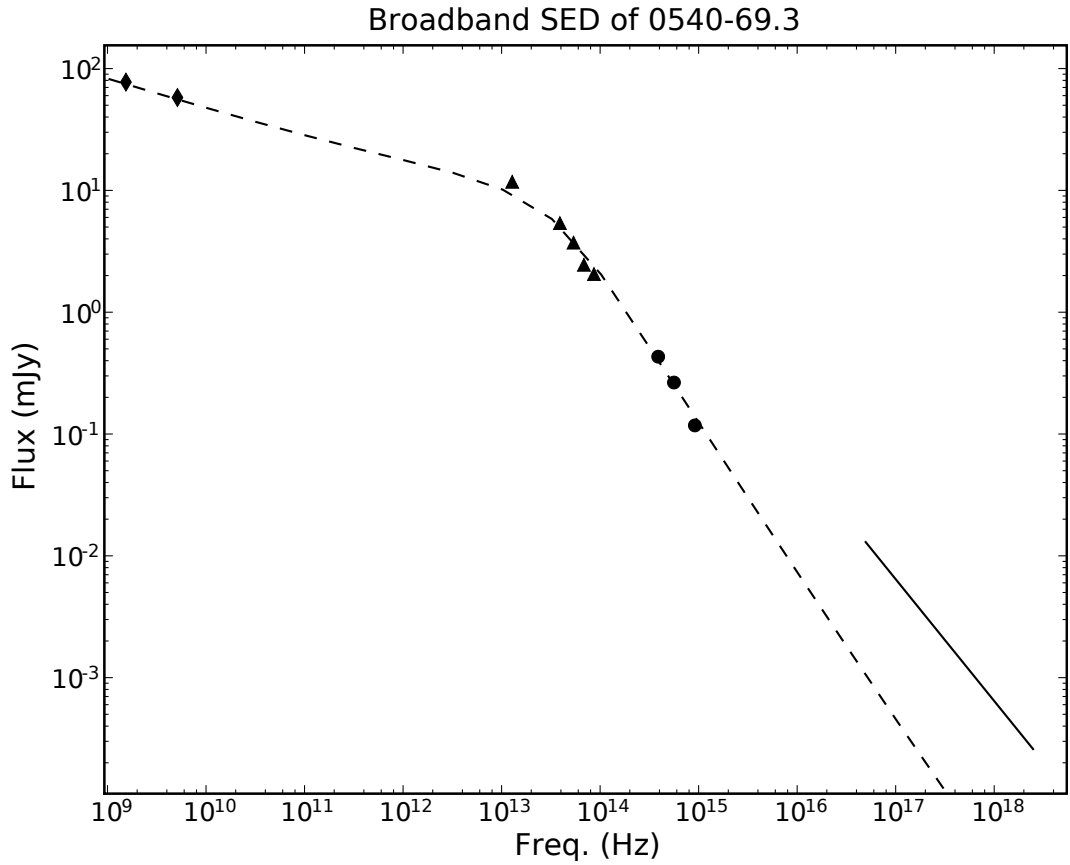}
\caption{Broadband spectrum of 0540.  Radio points (diamonds):
Manchester et al.~(1993).  IR points (triangles): our MIPS and IRAC fluxes.
Optical points (circles): Serafimovich et al.  ~(2004).  X-rays (solid line):
{\it Chandra} (Kaaret et al.~2001).  Dashed line: model described in
text.}
\label{bbspectrum}
\end{figure}

In order to subtract the synchrotron component, it was necessary to
produce a model synchrotron spectrum that includes the break in
power-law indices from optical to radio. The synchrotron model used
here is one of a class of simple outflow models in which various quantities
are allowed to have power-law dependencies on radius: flow-tube width,
flow velocity, gas density (where mass loading might allow a range of
possibilities), and magnetic-field strength (Reynolds, in
preparation).  Such models can produce synchrotron-loss steepening in
spectral index both steeper and flatter than the homogeneous-source
value 0.5 \citep{reynolds06}.  Here the model, used for illustrative
purposes, invokes a simple outflow geometry with conical flow tubes
(width $w \propto r$), mass increasing as radius (due presumably to
mass loading), flow velocity decreasing as $r^{-2}$ roughly, and
magnetic field as $r^{-1}$.  The initial magnetic field at the
injection radius is $B_0 = 2.5 \times 10^{-4}$ G.  This model predicts
a decrease in size with frequency as $\theta \propto \nu^{-0.34}$,
which might be slow enough to be consistent with observations,
especially as it might take place along the line of sight.  While this
is not meant as a definitive model for 0540, it describes the data
well as shown on Figure~\ref{bbspectrum} and was used to estimate the
synchrotron contribution. 

The result of radiative losses on electrons above the break energy in
a flat ($N(E) \propto E^{-s}$ with $s < 2$) energy distribution is for
such electrons to move to just below the break energy, where they can
produce a perceptible ``bump''.  However, the ``bump'' is almost
undetectable unless $s$ is very close to 0; for the 0540 value $s =
1.5$, there is almost no departure from the power-law below the break
frequency.  The model in Figure~\ref{bbspectrum} was calculated
including the redistribution of electron energies, and it can be seen
that the excess we observe below 24 $\mu$m cannot be attributed to
this cause.

\subsubsection{Fitting the Dust Component}

This left us with a residual rising continuum that we then fit with a
model dust spectrum. Since we presume that the dust present in 0540
would be newly formed ejecta dust, as seen in SN 1987A
\citep{ercolano07}, we have little {\it a priori} knowledge about the
grain-size distribution. However, since the wavelength of IR radiation
is much larger than typical ISM grain sizes, we adopt a model with a
single grain size, arbitrarily chosen to be $a=0.05$ $\mu$m in
radius. In any case, in the limit of $a$ $\ll \lambda$, the results
are independent of the choice of grain radius. We also do not know the
grain composition, as general results from the LMC should not apply to
ejecta dust. We thus consider two models; a graphite dust model and
the ``astronomical silicates'' model of \citet{draine84}. We calculate
the dust grain absorption cross section for both as a function of
wavelength. We then fit a simple modified blackbody model
(incorporating the grain absorption cross-section) to the data using a
least-squares algorithm designed for this model. We obtain a dust mass
of $\sim 3 \times 10^{-3}$ $\msun$ at a temperature of $50 \pm 8$ K
for silicate dust, while the resulting fit to the temperature with
graphite grains was slightly higher, $\sim 65$ K, and the required
dust mass was lower, $\sim 1 \times 10^{-3}$. 

The errors on the dust temperature are estimates based on using
different methods of removing lines and subtracting the background and
the underlying synchrotron spectrum. The resultant dust spectrum is
sensitive to these details. The dust mass should be considered
uncertain, and is probably only accurate to within a factor of a few,
as evidenced by the difference between derived masses for graphite and
silicate grains. Our data do not allow us to distinguish between
various dust compositions. It should also be noted that we are only
sensitive to dust that has been warmed by the shock wave from the
pulsar wind, and that there could be more dust that has not yet been
shocked, and is still too cool to be detected. Thus, our mass estimate
should be considered a lower limit.

\subsubsection{Grain Heating Mechanisms}
\label{grains}

We now turn our attention to heating mechanisms for this dust. We
first consider heating by the synchrotron radiation field from the
PWN. Since the spectrum of the synchrotron radiation is known in the
optical/UV portion of the spectrum and grain absorption cross-sections
can be calculated as a function of wavelength, it is possible to
estimate whether there is enough radiation to heat the dust to
temperatures observed. We calculate the optical depth of the dust
around the PWN, and integrate over all wavelengths from radio up 1
keV. Although the flux from the PWN is higher at longer wavelengths,
most of the absorption occurs in the optical/UV portion of the
spectrum, due to the steeply rising absorption cross-sections in this
regime. We compare this number to the luminosity in dust derived from
our dust model, $\sim 5 \times 10^{35}$ ergs s$^{-1}$. A simple
calculation showed that the radiation available falls short by several
orders of magnitude of what is necessary.

However, this method tells us nothing about the total amount of dust
that could be present to absorb the synchrotron radiation. Thus, to
further test this hypothesis, we calculated the temperature to which
dust would be heated if it were exposed to such an ultraviolet
radiation field. We find that dust would only be heated to $\sim{20}$
K. If this were the source of the emission seen in IRS, it would
predict a 70 $\mu$m flux that is several orders of magnitude higher
than the upper limit we have placed on emission there. Given that
these order of magnitude estimates are drastically inconsistent with
our observations, we consider heating by photons from the PWN to be
ruled out.

We then considered the somewhat more exotic possibility of the
observed excess arising from a protoplanetary disk around the pulsar,
unassociated with the nebula. It has long been known that planets can
form around pulsars \citep{wolszczan92}, and the supposition has been
that these planets arise from a protoplanetary disk around the pulsar,
the source of which has been attributed to several mechnisms
\citep{bryden06}. Various surveys of known pulsars have been made in
infrared and submillimeter wavelengths, but for the most part these
surveys have only produced upper limits on the dust emission present.

However, \citet{wang06} conducted a survey of neutron stars with IRAC
and found a debris disk around the young isolated neutron star 4U
0142+61. The authors suggest that the age of the debris disk compared
to the spin-down age of the pulsar favors a supernova fallback
origin. The IRAC observations combined with K-band Keck-I observations
suggest a multi-temperature thermal model with temperatures ranging
from 700-1200 K, where the disk has inner and outer radii of 2.9 and
9.7 $\rsun$, respectively. Using the same model the authors use
\citep{vrtilek90}, we calculate the necessary radius to reproduce
observed fluxes for 0540 for a disk with temperature $\sim{50}$ K to
be on the order of $10^{4}$ $\rsun$. A survey of disks around
Anomalous X-ray Pulsars (AXPs) \citep{durant05} found several
candidates for fallback disks which consistently had IR(K-band)/X-ray
flux ratios of order $10^{-4}$. Although we were not able to find any
archival near-infrared observations of the PWN, we can make an
estimate of this ratio by looking at the overall spectrum of the IRAC
and optical points. An estimate of $5 \times 10^{-2}$ is reasonable
for this ratio in 0540, significantly different than that found in the
AXPs.  Additionally, \citet{wang07} observed 3 known AXPs with
Spitzer, and found no mid-IR counterpart to any of them. Given these
discrepancies between these cases and that of 0540, we do not believe
that a protoplanetary disk around the pulsar is the origin of the
far-IR excess.

What then is the cause? Collisional heating by hot gas heated by
shocks driven into the ejecta can provide both a qualitative and
quantitative explanation for the dust present. Grain heating
rate, $\cal{H}$, goes as

\begin{equation}
\mathcal{H} \propto n_ev_{e}T_{e} \propto PT_{e}^{1/2},
\end{equation}
where $n_{e}$, $v_{e}$, and $T_{e}$ are electron density, velocity,
and temperature, and $P$ is the pressure, $nT$. In the PWN, P
is constant throughout the bubble, so that grain heating is more
efficient in hotter gas. We find that the slow, radiative shocks are
incapable of heating dust grains to temperatures much above $\sim 25$
K. Faster shocks, and thus higher temperatures, are required to heat
grains to observed temperatures.

To determine whether this is plausible, given the conditions in the
object, it is necessary to make an estimate of the amount of gas that
is still hot, i.e. capable of heating dust grains through collisions
with electrons. The shock cooling time \citep{mckee87} is

\begin{equation}
t_{cool} = 2.5 \times 10^{10} v^{3}_{s7}/\alpha \rho_{0},
\end{equation}
where $v_{s7}$ is the shock speed in units of $10^{7}$ cm s$^{-1}$,
$\rho_{0}$ is the pre-shock density in amu cm$^{-3}$, and $\alpha \ge
1$ is a parameter for the enhancement of cooling due to higher metal
content. We find that a shock with velocity $\sim 250$ km s$^{-1}$
would effectively heat dust to 50 K, with a pre-shock density of $\sim
8$ amu cm$^{-3}$, assuming the same pressure as in slow shocks. If the
dust component is composed of graphite grains at $\sim 65$ K, a
slightly faster shock of 325 km s$^{-1}$ is required. Using the above
equation, we find that the amount of hot gas is on the order of a few
tenths of a solar mass. This yields dust-to-gas ratios of a few
percent. Given the unknown dust content within the inner ejecta of a
supernova, we believe this is a reasonable explanation.

As a check on the constraints of such a fast shock, we calculated the
expected X-ray emission from such a shock, and found it to be below
the upper limits of thermal X-ray emission observed from the PWN,
except for very metal-rich ejecta.

\subsection{Origin of O-rich Clumps}
\label{clumps}

\cite{matzner99} considered a spherically-symmetric explosion of a 15
$M_\odot$ RSG, and found that its He core and heavy element ejecta
formed an approximately constant density, freely expanding ejecta
core.  C05 rescaled their results to other values of ejecta mass
$M_{ej}$ and kinetic energy $E_{51}$, arriving at the core density of

\begin{equation}
\rho_ct^{3} = 2.4 \times 10^{9} (M_{ej}/15 M_\odot)^{5/2}
E_{51}^{-3/2} {\rm g\ cm^{-3}\ s^{3}}.
\end{equation}
An additional compression is expected from the iron-nickel bubble
effect. For the centrally-located Ni with mass $M_{Ni}$, the adjacent
ejecta are expected to be swept up into a shell with velocity 

\begin{equation}
V_1 = 975 (M_{Ni}/0.1 M_\odot)^{1/5} (\rho_c t^{3}/10^{9} {\rm g\
cm}^{-3}\ {\rm s}^{3})^{-1/5} {\rm km\ s^{-1}}.
\end{equation}
The compression within the Fe-Ni bubble shell is at least by a factor
of 7, expected in strong, radiation dominated shocks with $\gamma =
4/3$. The shell density increases inward from a shock front to a
contact discontinuity separating the shocked ejecta from the Fe-Ni
bubble. In one dimensional hydrodynamical simulations, Wang (2005)
finds an average shell compression by a factor of 20. The average
shell density is then 

\begin{equation}
\rho_1 t^{3} = 4.8 \times 10^{10} (M_{ej}/15 M_\odot)^{5/2}
E_{51}^{-3/2} {\rm g\ cm^{-3}\ s^3}.
\end{equation}
(Diffusion of radiation might reduce this compression by a modest
factor of $\le 2$ -- Wang 2005.) At the current remnant's age of 1140
yr, the shell density is 

\begin{equation}
\rho_1 = 1.0 \times 10^{-21} (M_{ej}/15 M_\odot)^{5/2}
E_{51}^{-3/2} {\rm g\ cm^{-3}}.
\end{equation}
Because the dense iron-nickel bubble shell has been accelerated by
low-density gas within the bubble, the shell is subjected to the
Rayleigh-Taylor instability, and we expect it to fragment into
clumps. Within a factor of 2, their expected density is equal to the
preshock density for the O-rich clumps in 0540. We conclude that these
clumps are remnants of the iron-nickel bubble shell.

\cite{matzner99} found a sharp density drop by a factor of 10
at the interface between the He core and the H envelope, with an
approximately constant density through much of the H envelope. The
envelope density $\rho_{env}$ is then 

\begin{equation}
\rho_{env} t^{3} = 2.4 \times 10^{8} (M_{ej}/15 M_\odot)^{5/2}
E_{51}^{-3/2} {\rm g\ cm^{-3}\ s^3},
\end{equation}
200 times less dense than the iron-nickel bubble shell. This density
contrast is similar to the density contrast inferred between the
O-rich clumps and the more tenuous inter-clump gas. It is likely that
the PWN nebula expands now into the H envelope. Because the dense He
core has been decelerated by the less dense H envelope during the SN
explosion, the ensuing Rayleigh-Taylor instability led to a
large-scale macroscopic mixing between them. As a result, we expect a
two-phase medium ahead of the PWN shell, consisting of more tenuous
H-rich gas and denser He-rich gas. It is possible that shocks driven
into the He-rich gas became radiative; that could explain the presence
of H and He recombination lines in optical spectra of 0540.

The dense iron-nickel bubble shell should contain not only O-rich
ejecta, substantial amounts of He-rich gas are also expected. Slow (20
km s$^{-1}$) shocks driven into the dense He-rich gas may also become
radiative; if so, they could produce strong lines of low ionization
species. This could explain excess emission seen in optical and IR
spectra for low ionization species (see discussion in \S~4.2). More
detailed shock models are necessary to determine whether or not our
picture is consistent with observations.

The identification of dense O-rich clumps in 0540 with a compressed
and fragmented shell swept up by the iron-nickel bubble has important
implications for ejecta detection in SNRs. Dense O-rich clumps are
expected to produce strong optical or X-ray emission, once shocked and
heated by the reverse shock. The optical emission should be most
prominent for remnants with a particularly dense ambient medium,
either of circumstellar (e.g., Cas A) or interstellar (N132D)
origin. The entire class of optically emitting O-rich remnants may owe
its existence to the iron-nickel bubble effect. For ejecta expanding
into less dense ambient medium, X-ray emission is expected instead
since clumps will be reverse-shocked much later when their densities
dropped significantly because of free expansion. The O-rich clumps
such as seen in 0540, even when shocked to X-ray emitting temperatures
10,000 yr after the explosion, will have substantial ($\sim 1$
cm$^{-3}$) electron densities and emission measures. Even old remnants
should show O-rich ejecta in their interiors, in agreement with the
accumulating evidence gathered by modern X-ray satellites. A good
example is a 14,000 yr old SNR 0049 $-$73.6 in the SMC, where {\it
Chandra} imaging and spectroscopy revealed the presence of a clumpy
O-rich ring in its interior \citep{hendrick05}. Hendrick et
al. interpreted this ring as the shell swept up by the iron-nickel
bubble, based on mostly theoretical arguments. Observational evidence
for the iron-nickel bubble effect in 0540 strengthens this
interpretation for 0049$-$73.6, and possibly for many more mature SNRs
with detected ejecta emission in their interiors.

Dust formation is most likely to occur where ejecta density is the
highest. The dense O-rich clumps likely contain dust; this dust may
survive the passage through the radiative shock. If it were mixed into
the much hotter ambient medium, this surviving dust may be the source
of the observed infrared emission.

\section{Summary}
\label{concls}

We have observed the supernova remnant B0540-69.3 with all three
instruments aboard the {\it Spitzer Space Telescope}. We detected the
PWN in all 4 IRAC bands, as well as the 24 $\mu$m MIPS band. We did
not detect any emission from the PWN at 70 $\mu$m, though the upper
limit is rather unconstraining. We found no hint of detection at any
wavelength of the $\sim 30''$ shell surrounding the PWN, as seen in
X-rays and radio. Both the IRAC and the MIPS 24 $\mu$m photometric
fluxes are consistent with the emission being primarily dominated by
synchrotron emission, as synchrotron models extended both down from
the radio and up from optical wavelengths roughly reproduce the flux
seen in infrared. There is a change in slope of the overall
synchrotron spectrum taking place in mid-infrared wavelengths, as is
required to match the radio synchrotron power-law with the optical
power-law.

The IRS spectra in the 10-37 $\mu$m region show a clear excess of
infrared emission that cannot reasonably be attributed to any
synchrotron radiation. We conclude that this excess emission is coming
from a small amount ($\sim 1-3 \times 10^{-3} \msun$) of warm dust that
has been formed in the expanding ejecta from the SN. We consider
multiple heating mechanisms for this dust, ruling out both a fallback
disk around the neutron star and heating by the synchrotron radiation
from the PWN itself. We conclude that the dust is being heated by
shocks being driven into the ejecta by the energy input from the
pulsar. We derive a dust-to-gas mass ratio of the order of a few
percent, which is reasonable given how little is known about dust
content in the inner ejecta of SNe.

We consider the extended ($8''$ in radius) [O III] emission discovered
by Morse et al. in HST images of the nebula, and attribute this to
undecelerated ejecta that have been photoionized by photons from both
the radiative pulsar wind shocks and the synchrotron radiation from
the nebula. While there are not enough ionizing photons to do this
assuming solar abundances, we show that realistic assumptions about
the heavy element abundances in the ejecta, which are almost certainly
not solar, provide a plausible explanation of the [O III] halo.

We also detect a number of lines coming from both the ejecta in the
PWN and the background/foreground H II region. Most of the line
structures contained both a broad and a narrow component, blended
together due to the modest spectral resolution of the instrument. We
performed multi-gaussian fits to the line structures to identify both
of these components separately. The widths of the lines, as well as
their redshift from their rest wavelength, are broadly consistent with
previous line studies done in optical wavelengths. We find line widths
of order 1000-1300 km s$^{-1}$, and shifts between broad and narrow
components of lines of order 300-400 km s$^{-1}$. We model these
lines, as well as those found in optical wavelengths, and conclude
that slow ($\sim 20$ km s$^{-1}$) shocks driven into dense ($\sim 30$
cm$^{-3}$), O-rich clumps of material provide the most satisfactory
agreement with measured intensities. We find a preshock density
contrast of $\sim 100$ between the dense, optically-emitting clumps
and the rarefied gas behind the global shock, assuming rough pressure
balance between the two phases.

Our global picture of the pulsar-wind nebula consists of several
elements. An expanding, accelerating shell of material is driven into
the inner ejecta from the supernova, passing through the iron-nickel
bubble and the dense, clumpy shell, into which shocks are being driven
at 20 km s$^{-1}$. The fast, global shock has exited the bubble walls,
and has now reached a radius of about 1.2 pc. Beyond this shock, out
to a radius of 1.9 pc, material has been photoionized by UV photons
from both the shock and the synchrotron nebula, and this photoionized
material is observed in the form of an [O III] halo. This picture is
able to account for observations in the broad wavelength range from
radio to X-rays.

Future, high-resolution observations of this object in infrared
wavelengths, such as those which will be possible with the {\it James
Webb Space Telescope}, will serve to further its understanding. Just a
few of the possibilities that could be studied with such observations
are: spatial identification of the location of infrared lines, further
search for an infrared shell at $30''$ corresponding morphologically
with the X-ray shell, and identification of the spatial location of
the dust in the PWN. Further spectroscopy on the warm dust component
could potentially constrain the composition of dust formed out of
ejecta. The global shock is just one possible location for the hot gas
capable of heating dust grains, it is also possible that the shocked
and dusty heavy-element ejecta have been reheated in the turbulent and
hot PWN interior. The order of magnitude increase in the spatial
resolution of JWST can shed light on our hypothesis of the global
picture of the PWN. In addition, deep ground-based spectra of the [O
III] halo can confirm or refute the photoionization origin we have
suggested here.

\acknowledgments

We thank the referee for useful comments, and gratefully acknowledge
support through Spitzer Guest Observer grant RSA 170640.

\clearpage

\begin{deluxetable}{lc}

\tablecolumns{2}
\tablewidth{0pc}
\tabletypesize{\footnotesize}
\tablecaption{Measured Fluxes}
\tablehead{
\colhead{Channel} & Flux}

\startdata
IRAC Ch.1 (3.6 $\mu$m) & 1.77 $\pm${0.23}\\
IRAC Ch.2 (4.5 $\mu$m) & 2.19 $\pm${0.27}\\
IRAC Ch.3 (5.8 $\mu$m) & 3.61 $\pm${0.46}\\
IRAC Ch.4 (8.0 $\mu$m) & 5.10 $\pm${0.74}\\
MIPS Ch.1 (24 $\mu$m) & 13.19 $\pm${3.95}\\
MIPS Ch.2 (70 $\mu$m) & $<366$\\

\enddata

\tablenotetext{a}{All fluxes given in milliJanskys}

\label{fluxtable}
\end{deluxetable}

\def\res#1#2#3{$#1^{+#2}_{-#3}$}
\begin{deluxetable}{lcccccccc}
\tabletypesize{\scriptsize}
\rotate
\tablecaption{Line Fits\label{linetable}}
\tablewidth{0pt}
\tablehead{\colhead{}  & \multicolumn{3}{c}{Narrow Component} & \multicolumn{5}{c}{ Broad Component} \\

\colhead{Line} & $\lambda$ ($\mu$m) & Flux\tablenotemark{a} &
FWHM\tablenotemark{b} ($\mu$m) & $\lambda$ ($\mu$m) &
Flux\tablenotemark{a} & FWHM\tablenotemark{b} & FWHM (km s$^{-1}$) & Shift
(km s$^{-1}$)}

\startdata

[S IV] (10.5105) & 10.5165 & \res{2.12}{0.53}{0.53} & 1.75 & \res{10.5261}{0.0017}{0.0015} & \res{7.32}{1.0}{1.0} & \res{3.89}{0.32}{0.21} & \res{1110}{91}{60} & \res{+274}{49}{43} \\

[Ne II] (12.8135) & 12.8208 & \res{5.86}{0.50}{0.50} & 2.14 & \res{12.8436}{0.0036}{0.0034} & \res{4.98}{0.85}{0.85} & \res{4.28}{0.61}{0.66} & \res{1000}{72}{154} & \res{+534}{84}{80} \\

[Ne III] (15.5551) & 15.5639 & \res{4.59}{0.29}{0.29} & 2.59 & \res{15.5823}{0.0018}{0.0019} & \res{7.29}{0.56}{0.56} & \res{6.86}{0.31}{0.32} & \res{1320}{62}{62} & \res{+354}{35}{37} \\

[Fe II] (17.9359) & - & - & - & \res{17.9663}{0.0025}{0.0025} & \res{3.01}{0.38}{0.38} & \res{6.84}{0.54}{0.65} & \res{1140}{90}{109} & - \\

[S III] (18.7130) & 18.7236 & \res{2.22}{0.40}{0.40} & 3.12 & \res{18.7407}{0.0012}{0.0011} & \res{10.18}{0.59}{0.59} & \res{6.07}{0.18}{0.16} & \res{972}{28}{26} & \res{+274}{19}{19} \\

[O IV] (25.8903) & - & - & - & \res{25.9454}{0.025}{0.0062} & \res{5.32}{1.7}{1.7} & \res{13.39}{3.8}{1.4} & \res{1650}{300}{180} & - \\

[Fe II] (25.9883) & - & - & - & 26.0375 & \res{1.71}{0.61}{1.7} & 10.03 & 1140 & - \\

[Si II] (34.8152) & 34.8419 & \res{5.13}{0.27}{0.27} & 5.81 & \res{34.8875}{0.0018}{0.0061} & \res{2.75}{0.31}{0.31} & \res{8.42}{0.25}{0.79} & \res{724}{22}{68} & \res{393}{16}{53} \\

\enddata

\tablenotetext{a}{Flux in units of 10$^{-14}$ ergs cm$^{-2}$ s$^{-1}$}
\tablenotetext{b}{FWHM in units of 10$^{-2} \mu$m}

\tablecomments{Centroid position and FWHM of narrow components fixed
to values specified in the text. [Fe II] at 26 $\mu$m fixed to
redshift and FWHM of [Fe II] 17.9 $\mu$m. Col. (9): Shift of broad
line relative to narrow line.}

\end{deluxetable}

\begin{table}
\begin{center}
\centerline{Table 3}

\vspace*{2mm}
\centerline{Normalized Emission Line Fluxes}

\vspace{4mm}
\begin{tabular}{| lrrrr |}

\hline \hline
Line            & M06  & K89   & Spitzer & Model \\
\hline
 O II  3727     & 46.  & 39.   & - & 52.8  \\ 
 Ne III  3869   &  7.2 & $<$1.5 & - &  9.3 \\
 S II    4072   &  3.6 &  3.   & - & 4.9  \\
 O III   4363   &  3.3 &  7.   & - & 4.2 \\
 O III   5007   & 100. & 100.  & - & 100. \\
 Fe VII  6085   &  -   &  2.:  & - & 0.02 \\
 O I     6303   &  3.3 &  5.   & - & 0.9 \\
 S II    6722   & 33.8 & 67.   & - & 36.1 \\
 Ar III  7136   &  -   &  8.   & - & 19.2 \\
 Ca II   7291   &  -   &$<$2.  & - & 0.6 \\
 O II    7325   &  -   &  6.   & - & 3.6 \\
 Fe II   8617   &  -   &  2.   & - & 5.7 \\
 S III   9532   &  -   &  34.  & - & 30.0 \\
 O I     7774   &  -   &  -    & - & 0.01 \\
 S IV    10$\mu$m&  -   &  -    & 26. & 7.4 \\
 Ne II   12$\mu$m&  -   &  -    & 31. & 3.2 \\
 Ne III  15$\mu$m&  -   &  -    & 33. & 29.0 \\
 Fe II   17$\mu$m&  -   &  -    &  8.4&  7.6 \\
 S III   18$\mu$m&  -   &  -    & 35. & 20.5 \\
 O IV    26$\mu$m&  -   &  -    & 15. & 11.6 \\
 Fe II   26$\mu$m&  -   &  -    &  4.8& 12.8 \\
 Si II   35$\mu$m&  -   &  -    & 22. & 7.9 \\
\hline

\end{tabular}
\label{linemodel}
\end{center}

\end{table}

\clearpage

\appendix

\section{PHOTOIONIZATION CALCULATION}

There are two sources of ionizing photons that can pre-ionize the
material ahead of the shock; ionizing photons produced behind
radiative shocks, and those produced by relativistic electrons in the
form of synchrotron radiation. We examine each of these in
turn. Detailed photoionization calculations would require modeling
that is beyond the scope of this paper, and we present calculations
that are only intended to be rough estimates. Since we do not have a
detailed, multi-dimensional model that provides the shock dynamics
after it encounters the iron-nickel bubble, we here detail the
calculations done in the absence of the bubble, assuming the models of
C05 describe the global shock encountering the inner ejecta. We intend
only for this rough calculation to show that photoionization is a
plausible mechanism for ionizing material out to $8''$.

First, it is necessary to determine the amount of ionizing radiation
emergent from behind the shock. \citet{shull79} give emergent photon
number fluxes per incoming hydrogen atom as a function of shock
speed. Since we know both the density and the shock speed in 0540 as a
function of time for a given density profile, we are able to calculate
the number of ionizing photons emerging from the shock over the
lifetime of the remnant. Here we consider the $m=1.06$ case. We count
all photons with energies above 13.6 eV as ionizing. However, Shull \&
McKee only considered shocks up to 130 km s$^{-1}$. By calculating the
cooling time from equation (2), we see that shocks in 0540 are
radiative up to speeds of over 150 km s$^{-1}$. In order to
extrapolate the numbers given in Shull \& McKee, we use figure 13 of
\citet{pun02}, and assume a single constant factor as the relationship
between the total number of H$\alpha$ photons and the total number of
ionizing photons. We then simply integrate the total number of
ionizing photons throughout the lifetime of the nebula.  We exclude
early times when densities were high enough that recombination times
were shorter than the age of the remnant (about the first 450
years). Using the same conditions as were used above for modeling the
nebula, we find that photoionization from radiative shocks can ionize
0.53 $\msun$.

Next, we calculated the ionizing flux from the synchrotron nebula
itself. We used the optically determined synchrotron power-law of
$\alpha = -1.1$, and considered photons from the Lyman alpha limit up
to 1 keV, though the choice of the upper limit has little effect due
to the steep drop of the synchrotron spectrum. In order to integrate
the luminosity of the nebula over time, it was necessary to use the
time evolution power-law index of $l=0.325$ \citep{reynolds84}. We
considered the emission from the nebula from after the time that
recombinations were important up through the presumed age of the
remnant (450-1140 yrs.) We find enough ionizing photons to ionize 0.21
$\msun$.

We then calculated how far out the ionization front would extend, to
see if this could account for the [O III] emission at $8''$ observed
by Morse et al. Using $m=1.06$ as the power-law index for the ejecta
density profile, the relation between the mass and radius of the
ionization front to that of the shock front can be written as

\begin{equation}
({M_{if}\over M_{sh}})^{0.515} = {R_{if}\over R_{sh}},
\end{equation}

where $M_{if}$ is the mass ionized by both mechanisms, plus the mass
swept up during the early stages of the remnant when recombinations
were occuring, and $M_{sh}$ is the mass swept by the shock, given
above as 0.75 $\msun$. With $M_{if} = 1.25 \msun$ and $M_{sh} = 0.75
\msun$, we find a ratio of $R_{if}$ to $R_{sh}$ of 1.3. While this is
not quite enough to account for the observed [O III] emission at 1.8
pc, this is almost certainly an underestimate of the amount of
photoionized material.

The same calculations for the case of a flat density profile yield the
following values. UV photons from the radiative shocks can photoionize
0.83 $\msun$, while the synchrotron photons from the nebula can ionize
0.18 $\msun$ (the difference in this number is due to the fact that
the recombination timescale is slightly longer for the higher
densities involved in this case, thus fewer photons are included in
the final photon count). The shock itself sweeps up 0.95 $\msun$, and
the relation between the mass interior to the ionization front and the
mass interior to the shock front is given by 

\begin{equation}
({M_{if}\over M_{sh}})^{1/3} ={R_{if}\over R_{sh}}. 
\end{equation}

We find that the ionization front is 1.2 times
farther out than the shock front. Again, while this is not enough to
account for what is observed, it can be considered a lower limit.

As a possible resolution to this, we return to the issue of heavy
element abundances. The calculations above assume standard solar
abundances, but, one would clearly expect the shock encountering the
ejecta to be overtaking material that is higher in metallicity than
solar. If the ejecta that the shock is running into is enriched in
helium and other heavier elements, more mass can be ionized per
ionizing photon (differences in ionization potential
notwithstanding). Since, for the case of $m=1.06$, a modest factor of
${\sim 2}$ in the amount of shock ionized mass would account for the
emission seen at $8''$, this is an entirely plausible explanation.

\section{SPHERICAL MODEL}

We include here the results from our spherically symmetric
model. Although these results indicate that such a model is not able to
account for line emission, it was nonetheless an important starting
point for our more complete models.

In order to model the PWN, it is necessary to determine the inner
ejecta density profile. \cite{matzner99} examined the relationship
between the progenitor of a core-collapse supernova and the resulting
density distribution of the ejecta. They find that core-collapse SNe
lead to density profiles that are best fit by two components, an inner
component that is relatively flat, and an outer component that is
extremely steeply dropping. In the case of a red supergiant (RSG), the
flat inner ejecta correspond to the mass contained in the helium core
of the progenitor star, a few solar masses of material. In
approximating these results for the cases of type Ib/c and type IIP
supernovae, C05 uses the expression $\rho_{SN} = At^{-3}(r/t)^{-m}$,
where $m=0.0$ and $1.06$ for the inner ejecta of type IIP and type
Ib/c SNe, respectively. He concludes that 0540 is the result of an
explosion of a Wolf-Rayet star, and thus should have little or no H in
the inner ejecta. However, in light of recent optical observations
that have detected H lines in the inner ejecta
\citep{serafimovich04,morse06}, it is now believed \citep{chevalier06}
that 0540 is a type IIP, the result of a red supergiant.

However, the power-law approximations of C05 do not take into account
any mixing of ejecta. Even if the progenitor star did explode as a
type IIP, any mixing of ejecta would steepen the power-law index from
a flat distribution to one that declines as a function of radius. We
therefore consider values of $m$ of both 0 and 1.06 here.

We assume the standard picture of a pulsar emitting magnetic-dipole
radiation at the spin frequency, slowing down with a constant
braking index, $n$, defined by ${\dot \Omega}
\propto -\Omega^{n}$.  Then the total pulsar energy loss ${\dot E}(t)$
is given by

\begin{equation}
{\dot E(t)} = {{\dot E}_0 \over (1 + {t\over \tau})^{(n+1)/(n-1)}}
\label{edot}
\end{equation}

where $\tau$ is a slowdown timescale related to the characteristic
time $t_{\rm ch} \equiv P/2{\dot P}$ by 

\begin{equation}
\tau = {2t_{\rm ch}\over n-1} -t.
\end{equation}

Several different values for the braking index have been reported in
recent years; we adopt the most recent measurement of $n=2.14$
\citep{livingstone05}. Assuming an age of $t= 1140$ yr, $P=50$ ms,
${\dot P}=4.8 \times 10^{-13}$ s s$^{-1}$, and characteristic time
$t_{\rm ch} = 1655$ yr, we find $\tau = 1770$ yr.  We assume a current
pulsar spindown energy input of ${\dot E}=1.5 \times 10^{38}$ ergs
s$^{-1}$. From this we can calculate ${\dot E_{0}}$ according to
Equation~\ref{edot}.

Using the X-ray determined radius of $5''$, or approximately 1.2 pc,
we apply the model of C05 for the accelerating PWN bubble driven into
the cold ejecta. We first consider a model with a perfectly flat inner
ejecta density profile, i.e. $m=0$. The model yields a shell velocity
$V_{\rm shell}$ that is currently 1170 km s$^{-1}$, with a shock
velocity $V_{\rm shock}$ (that is, the difference in the shell
velocity and the free-expansion velocity of the ejecta) of 150 km
s$^{-1}$. The current pre-shock density of the ejecta, $\rho_{0}$, is
$9.2 \times 10^{-24}$ g cm$^{-3}$, and the shock has swept up a total
mass in gas, $M_{\rm swept},$ of 0.95 $\msun$. For the $m=1.06$ case,
we find a somewhat higher shell and shock velocity, as would be
expected since the shell is encountering less dense material as it
expands, relative to $m=0$. We find $V_{\rm shell} = 1200$ km s$^{-1}$
and $V_{\rm shock} = 190$ km s$^{-1}$, with $\rho_{0} = 4.7 \times
10^{-24}$ g cm$^{-3}$ and $M_{swept} = 0.75$ $\msun$.

As we will show, the data favor the case of $m=1.06$, and in fact
argue for an even steeper density profile. A flat distribution would
overpredict certain optical lines, as discussed below. In addition, we
show in Appendix B a rough estimate of the amount of ionizing
radiation available (both thermal and synchrotron) 
to produce the [O III] halo seen out to $8''$. 
For the case of $m=0$, we need nearly 5 times more ionizing
photons to account for the material seen at $8''$. For $m=1.06$, we
only need a factor of $\sim 2$. 
While our estimates are probably only good to a factor of 2, the models
clearly prefer steeper density profiles.

Line strengths can also help distinguish between ejecta density
profiles. \citet{chevalier92} investigated the cooling time of the
post-shock gas in an SNR. For the case of $m=0.0$, the density ahead
of the shock is high enough that the cooling times for the remnant are
short compared with the age of the remnant. Enhancements in heavy
element abundances shorten the cooling times further. Because of this,
the shock quickly becomes radiative, and a fast ($\sim 150$ km
s$^{-1}$), radiative shock will significantly overpredict several
lines, including [O III] and [Fe VII]. It is possible that Fe is over
abundant, but then the observed [Fe II] IR line would have to come
from somewhere else.

As a resolution to this problem, we explore the effect of different
density profiles on the power radiated in lines behind the shock from
shocked gas in the process of cooling. Assuming \citep[as
in][]{mckee87} that the cooling curves of \cite{raymond76} can be
approximated as $\Lambda \propto T^{-1/2}$, we use the following
expression for the radiated power from the cooling layer behind the
shock:

\begin{equation}
P \propto \int_{shell} \rho_e \rho_H \Lambda(T) dV.
\end{equation}

Since the models of C05 give the density of material entering the
shock, we were able to numerically integrate the radiated power over
the thickness of the cooling layer, where we define the limits of
integration of the cooling layer as the thickness of the layer in
which the gas cools from its immediate post-shock temperature down to
10$^{4}$ K. In terms of relative power, the $m=1.06$ model radiated
about 45\% less power. We also ran a model with $m=2.0$, and found a
factor of about 3.5 less energy radiated. We do not use this model to
favor a particular value of $m$, only to demonstrate that any mixing of
the inner ejecta, which would likely lead to a value of $m$ for the
average density greater than 0, would reduce the amount of emission
radiated in lines. 

The spherically symmetric model is thus insufficient to describe the
data in two ways. Densities are not high enough to account for
observed optical and IR lines, and fast radiative shocks would
overpredict lines that are not seen, such as [O III] and [Fe VII]. Our
model discussed in the main text provides a potential solution to both
problems in the form of an iron-nickel bubble in the inner
ejecta. Because the fast shock initially propagated through the
low-density medium of the bubble, [O III] and [Fe VII] lines should
not be strong, and the passage of the shock through the high-density
bubble wall would provide the dense environment necessary for lines
that are observed.

\end{document}